\documentclass[preprint2]{aastex}

\slugcomment{Accepted to the Astrophysical Journal}

\shorttitle{Atmospheric Circulation of GJ436b}
\shortauthors{Lewis et al.}

\def\tt{\bf}

\pdfoutput=1

\begin{document}

\title{Atmospheric Circulation of Eccentric Hot Neptune GJ436b}

\author{Nikole K. Lewis and Adam P. Showman}
\affil{Department of Planetary Sciences and Lunar and Planetary Laboratory, 
The University of Arizona, Tucson, AZ 85721}
\email{nlewis@lpl.arizona.edu}

\author{Jonathan J. Fortney}
\affil{Department of Astronomy \& Astrophysics, 
University of California, Santa Cruz, CA 95064}

\author{Mark S. Marley and Richard S. Freedman\altaffilmark{1}}
\affil{NASA Ames Research Center 245-3, Moffett Field, CA 94035}

\and

\author{Katharina Lodders}
\affil{Washington University, St. Louis, MO 63130}

\altaffiltext{1}{SETI Institute, 515 N. Wishman Road, Mountain View, CA 94043}

\begin{abstract}
GJ436b is a unique member of the transiting extrasolar planet population being one of the smallest and least 
irradiated and possessing an eccentric orbit. Because of its size, mass and density, GJ436b could plausibly have 
an atmospheric metallicity similar to Neptune (20-60 times solar abundances),  which makes it an ideal target to 
study the effects of atmospheric metallicity on dynamics and radiative transfer in an extrasolar planetary atmosphere.  We 
present three-dimensional atmospheric circulation models that include realistic non-gray radiative transfer for 1, 3, 10, 30, and 
50 times solar atmospheric metallicity cases of GJ436b.  Low metallicity models (1 and 3 times solar) show little 
day/night temperature variation and strong high-latitude jets.   In contrast, higher metallicity models 
(30 and 50 times solar) exhibit day/night temperature variations and a strong equatorial jet.    
Spectra and light curves produced from these simulations show strong orbital phase dependencies in the 50 times solar 
case and negligible variations with orbital phase in the 1 times solar case.  Comparisons between the predicted planet/star 
flux ratio from these models and current secondary eclipse measurements support a high metallicity 
atmosphere (30-50 times solar abundances) with disequilibrium carbon chemistry at play for GJ436b.   
Regardless of the actual atmospheric composition of GJ436b, our models serve to illuminate how metallicity influences the 
atmospheric circulation for a broad range of warm extrasolar planets.

\end{abstract}


\section{Introduction}

The ``hot Neptune'' GJ436b was first discovered by \citet{but04} and later determined to transit 
its host star as seen from earth by \citet{gil07a}. Since the discovery of its transit and subsequent secondary 
eclipse, GJ436b has become a popular target for {\it Hubble} \citep{bea08,pon09} and {\it Spitzer} \citep{gil07b, demi07, demo07,ste10} 
observations as well as modeling efforts \citep{spi10,mad10}. Given GJ436b's mass 
($M_p$=0.0729M$_J$) and radius ($R_p$=0.3767R$_J$), its interior must contain significant quantities of heavy elements
in addition to hydrogen and helium \citep{ada08, fig09, rog10, net10}.
This raises the possibility that, like Uranus and Neptune, whose atmospheric C/H ratios lie between 20 and 60 times solar \citep{gau95},
the atmosphere of GJ436b is highly enriched in heavy elements.  This makes GJ436b an excellent case study for 
atmospheric chemistry, radiative transfer, and global circulation that should differ significantly from the well 
studied ``hot Jupiters'' HD209458b and HD189733b.

Observations of HD189733b using the {\it Spitzer Space Telescope} provided the first clear evidence for atmospheric circulation 
on an extrasolar planet \citep{knu07, knu09}.  Most efforts to model atmospheric circulation for extrasolar planets have focused on 
hot Jupiters, specifically HD189733b and HD209458b \citep{sho02, sho08, sho09, cho03, cho08, coo05, coo06, dob08, men09, rau10}.  Only 
a handful of studies have specifically investigated the effects of non-synchronous rotation \citep{cho08, sho09}, non-zero 
obliquity \citep{lan07}, and non-zero eccentricity \citep{lan08}.  The possible effect of atmospheric composition, and hence opacity, 
on circulation patterns that may develop on extrasolar planets has been investigated to some extent by \citet{dob08} and \citet{sho09}, 
but largely ignored in most of the current two-dimensional and three-dimensional atmospheric models.  Atmospheric composition is key in 
determining opacity and radiative timescales that play a crucial role in the development of circulation on these planets.  

Here we present three-dimensional atmospheric models for GJ436b that
incorporate both equilibrium chemistry and realistic non-gray radiative transfer.    Although the actual
composition of GJ436b's atmosphere is likely to deviate from an equilibrium chemistry solution as shown from the
secondary eclipse observations of \citet{ste10}, our investigation still serves to explore the effect metallicity can
play in controlling the atmospheric circulation not only on GJ436b but for a broad range of gaseous extrasolar planets
in a similar temperature range.  Our models are not constrained to match measured chemical abundances and
temperatures, but instead provide a systematic look at how changes in atmospheric metallicity over the range from 1 times to 50
times solar values affects the basic thermal and dynamical structure of the planet's atmosphere.  We do not expect that
spectra and light curves from our model will provide a match to observational data, but instead illuminate some of
the underlying atmospheric physics responsible for current observations and suggest areas of focus for future
observations.  Section~\ref{model} gives an overview of the three-dimensional coupled radiative transfer and atmospheric
dynamics model used in this study.   Section~\ref{results} presents the global thermal structures and winds that develop in each
of our models along with predicted light curves and emission spectra.  Sections~\ref{dis} and~\ref{conc} provide a
brief discussion of the results and final conclusions.

\section{Model}\label{model}

The atmospheric model used in this study is a three-dimensional (3D) coupled radiative transfer and dynamics model that was specifically developed with 
the study of extrasolar planetary atmospheres in mind.  The Substellar and Planetary Atmospheric Radiation and Circulation (SPARC) model is described 
in detail in \citet{sho09} as applied to HD189733b and HD209458b.  A basic overview of the SPARC model along with the specific changes made to the 
model setup for GJ436b are presented here for completeness.  The SPARC model employs the MITgcm \citep{adc04} to treat the atmospheric 
dynamics using the primitive equations, which are valid in stably stratified atmospheres where the horizontal dimensions of the 
flow greatly exceed the vertical dimension.  
For GJ436b, the horizontal length scale of the flow is $\sim10^{7}$ m while the vertical scale height of the atmosphere is $\sim300$ km.  The simulations 
presented here take advantage of the cubed-sphere grid \citep{adc04} at a resolution of C32 (roughly $64\times128$ in latitude and longitude) to 
solve the relevant dynamic and energy equations.   The vertical dimension in these simulations spans the pressure ($p$) range from 200 bar to 20 $\mu$bar 
with 47 vertical levels, evenly spaced in $\log(p)$.  The boundary conditions in our simulations are an impermeable surface at the bottom and a zero 
pressure surface at the top both of which are free slip in horizontal velocity.  

We have coupled the MITgcm to the non-gray radiative transfer model of \citet{mar99} to realistically determine the magnitude of 
heating/cooling at each grid point.   The radiative transfer model, a two-stream version of the \citet{mar99} plane-parallel code, 
assumes local thermodynamic equilibrium and includes intensities over the wavelength range from 0.26 to 300 $\mu$m.  The opacity at each 
pressure-temperature-wavelength grid point is tabulated using the correlated-k method \citep{goo89}.  Our extensive opacity database is 
described in \citet{fre08}.  The chemical mixing ratios, which are computed assuming thermochemical equilibrium, are calculated as in 
\citet{lod02, lod06}.  Calculated opacities assume a gaseous composition without particulate matter and account for the possibility of 
chemical rainout.  Because GJ436b plausibly has an atmospheric chemistry that is enhanced in heavy elements, we developed opacity 
tables for 3 times (3$\times$), 10 times (10$\times$), 30 times (30$\times$), and 50 times (50$\times$) solar metallicity in 
addition to the 1 times (1$\times$) solar metallicity opacity table.  In the enhanced metallicity opacity tables, all elements other than 
hydrogen and helium are assumed to be enhanced by the same factor over current solar values.  The opacity databases of \citet{fre08} were
updated to include the opacity effects of CO$_2$, which is an important carbon bearing species at higher metallicities.  The full opacity 
tables are divided into 30 wavelength bins as outlined in \citet{sho09}.  This binning of opacities allows for greater computational efficiency
while only introducing small ($<1\%$) deviations from the net radiative flux calculated with higher resolution opacity tables.   

\begin{figure}
\centering
\includegraphics[width=0.49\textwidth]{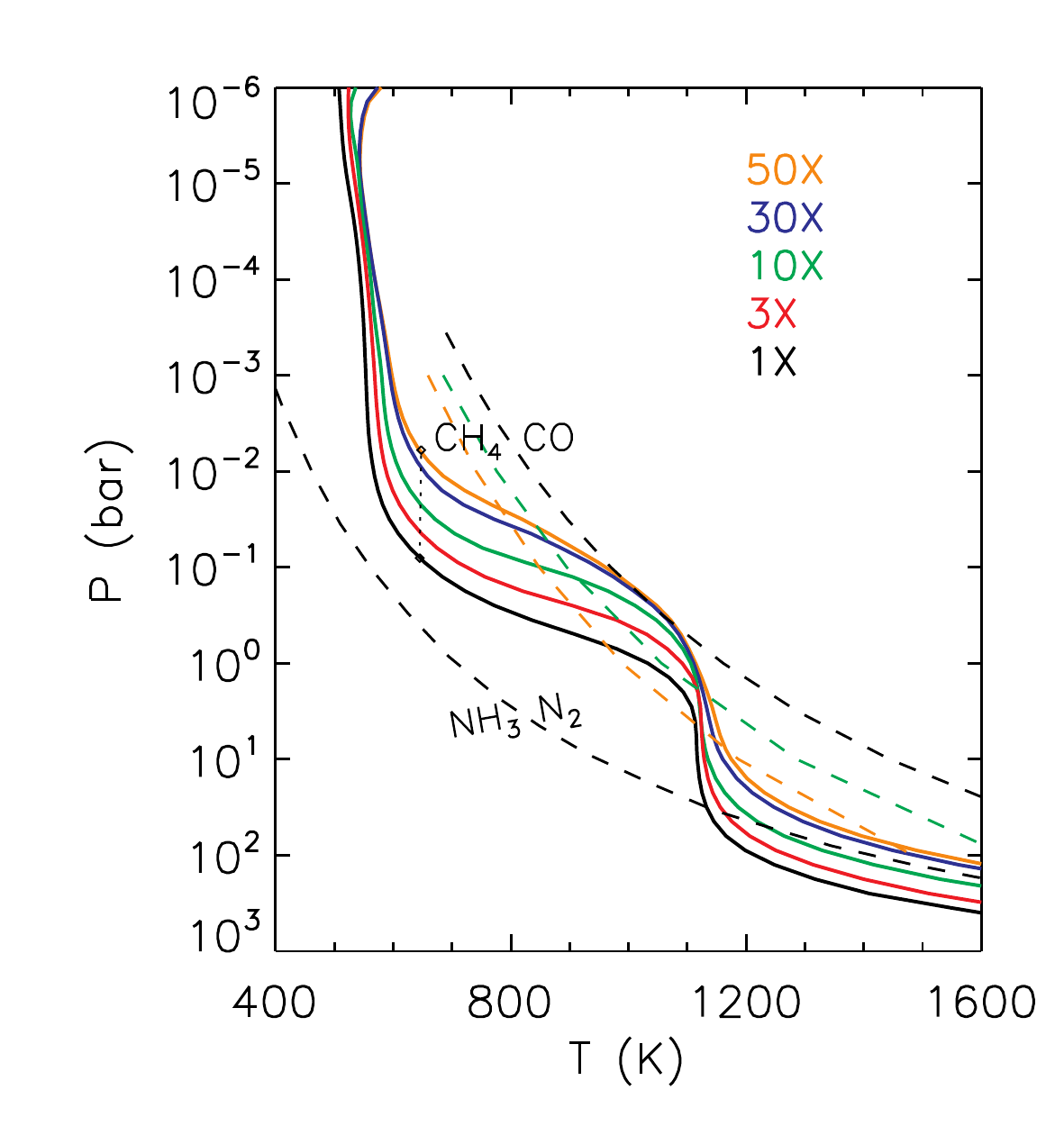}
\caption{Initial pressure temperature profiles assumed for each metallicity case of GJ436b.  Lines of equal abundance for CH$_4$ vs CO 
and NH$_3$ vs N$_2$ are shown to highlight the dominant carbon and nitrogen bearing species at each pressure level for each 
metallicity case.  As the metallicity in the atmosphere is increased form 1$\times$ to 50$\times$ solar, the dominant carbon bearing 
species changes from CH$_4$ to CO.  Diamonds represent the level of the mean photosphere ($T=T_{eff}$), which decreases in 
pressure as the metallicity is increased.  Profiles assume planet-wide redistribution of absorbed incident energy.
(a color version of this figure is available in the online journal.)}\label{chem_plot}
\end{figure}

For each model, the winds are assumed to initially be zero everywhere and each column of the grid is assigned the same pressure-temperature 
profile.  This initial pressure-temperature profile is derived from one-dimensional radiative-equilibrium calculations performed using the 
radiative transfer code in the absence of dynamics.  Figure~\ref{chem_plot} shows the pressure-temperature profiles derived for each metallicity 
case of GJ436b.  These pressure-temperature profiles were derived using the methodology presented in \citet{for08, for05}.   
The physical properties assumed for GJ436b and its host star (GJ436A) are presented in Table~\ref{gj436_params}.  Using these planetary 
and stellar parameters, the effective temperature ($T_{eff}$) of GJ436b is calculated to be 649 K, assuming planet-wide redistribution of the 
incoming stellar flux.  This $T_{eff}$ corresponds to a mean photospheric level\footnote{Defined in this context as the atmospheric pressure
where the local temperature equals the effective temperature.} of 1 to 100 mbar depending on the assumed metallicity of the atmosphere 
(Figure~\ref{chem_plot}).  
  
Because GJ436b is known to have an eccentric orbit, we incorporated the effects of non-synchronous rotation and time-varying distance from the host 
star into the SPARC model.  The most probable rotation rate for GJ436b was determined using the following pseudo-synchronous  
rotation relationship presented in \citet{hut81}:
\begin{equation}
P_{rot}=P_{orb}\left[\frac{(1+3e^2+\frac{3}{8}e^4)(1-e^2)^{3/2}}{1+\frac{15}{2}e^2+\frac{45}{8}e^4+\frac{5}{16}e^6}\right]
\end{equation}
where $P_{rot}$ is the planetary rotation rate, $P_{orb}$ is the orbital period of the planet, and $e$ is the eccentricity of the planetary 
orbit.  In all cases considered here the obliquity of the planet is assumed to be zero.  The time-varying distance of the planet with 
respect to its host star, $r(t)$, is determined using Kepler's equation \citep{mur99} and used to update the incident flux on the planet 
at each radiative timestep.  A diagram of GJ43b's orbit is presented in Figure~\ref{orbit_fig}.  To test the impact of 
pseudo-synchronous rotation and time-varying stellar insolation, additional simulations for the 1$\times$ and 30$\times$ solar metallicity 
cases were performed assuming synchronous rotation and zero eccentricity. 

In our models, for computational efficiency, the radiative timestep used to update the radiative fluxes is longer than 
the timestep used to update the dynamics.   Generally, as we increased the metallicity of the atmosphere, progressively shorter 
radiative and dynamical timesteps were needed to maintain stability.  For the 1$\times$ and 3$\times$ solar metallicity cases a dynamic 
timestep of 25 s and a radiative timestep of 200 s were used.   The 10$\times$ and 30$\times$ solar metallicity cases required a 
dynamic timestep of 20 s and a radiative timestep of 100 s while the 50$\times$ solar case required a dynamic timestep of 15 s and 
a radiative timestep of 60 s.  Timestepping in our simulations is accomplished through a third-order Adams-Bashforth scheme \citep{dur91}.  
We applied a fourth-order Shapiro filter in the horizontal direction to both velocity components and the potential temperature
over a timescale equivalent to twice the dynamical timestep in order to reduce small scale grid noise while minimally affecting the 
physical structure of the wind and temperature fields at the large scale.  

We integrated each of our models until the velocities reached a stable configuration.  Figure~\ref{vrms_plot}
show the root mean square (RMS) velocity as a function of pressure and simulated time, calculated according to:
\begin{equation}
V_{RMS}(p)=\sqrt{\frac{\int (u^2+v^2)\,dA}{A}}
\end{equation}
where the integral is a global (horizontal) integral over the globe, $A$ is the horizontal area of the globe, 
$u$ is the east-west wind speed, and $v$ is the north-south wind speed.  The high-frequency variations in the 
RMS velocity seen in the upper levels of both the 1$\times$ and 50$\times$ solar cases are largely due to variation in 
the incident stellar flux associated with the eccentric orbit of GJ436b.  Notice that, in the observable atmosphere 
(pressures less than 100 mbar), the orbit-averaged winds become essentially steady
within $\sim$2500 Earth days for solar metallicity and $\sim$1000 Earth days for 50$\times$ solar metallicity.  RMS wind speeds typically
reach $\sim$1 km s$^{-1}$ at photosphere levels. Any further increases in wind speeds will be small and confined to pressure well 
below the mean photosphere so as not to affect any synthetic observations derived from our simulations.  
As outlined in \citet{sho09} the energy available for the production of winds is limited largely by the global available potential 
energy within the atmosphere and to some extent energy losses due to the Shapiro filter which acts as a hyperviscosity.  A full discussion 
of the energetics of our simulated GJ436b-like atmosphere is left for a future paper.                  
   
\section{Results}\label{results}

The following sections overview the key results from the study of GJ436b's atmospheric circulation at 1$\times$, 3$\times$, 
10$\times$, 30$\times$, and 50$\times$ solar metallicity.  Both the thermal structure and winds in these simulations 
have a strong dependence on the assumed composition of the atmosphere for GJ436b.  Additionally, theoretical light curves 
and spectra are produced from our 3D model atmospheres and compared with available data.  

\subsection{Thermal Structure and Winds: Dependence on Metallicity}\label{metallicity}

Figure~\ref{uvt_plot} presents snapshots of the temperature and wind fields at three pressure levels in the atmosphere for 
the 1$\times$ and 30$\times$ solar cases near secondary eclipse when the full day-side of the planet faces Earth (Figure~\ref{orbit_fig}).  
Overall, the 30$\times$ solar case is significantly ($\sim100$ K) warmer than the 1$\times$ solar case at each pressure.  
The increased atmospheric opacity that comes with metallicity enhancements leads to an upward shift in the pressure-temperature 
profiles. This effect is self-consistently generated in the three-dimensional model integrations but can
also be seen in the one-dimensional radiative-equilibrium solutions shown in Figure~\ref{chem_plot}.  Overall, the day/night temperature contrast in the 
upper layers of the atmosphere ($\sim$1 mbar) and the equator/pole temperature contrast deeper in the atmosphere ($\sim$30 mbar) 
increase with atmospheric metallicity.   However, because the pressure at a given optical depth is smaller at high metallicity than low 
metallicity, the regions that develop significant day/night temperature contrasts shifts to higher altitude as metallicity increases.   
At the 1 bar level, the equator/pole temperature contrast in the 30$\times$ solar case is smaller than that in the 1$\times$ solar case.  
This lack of a strong temperature contrast at 1 bar in the 30$\times$ solar case occurs because this pressure is at a greater optical depth 
in the 30$\times$ solar case than in the 1$\times$ solar case, and thus occurs below the levels with the strongest heating/cooling. 

It is also informative to compare the flow patterns indicated by the arrows in Figure~\ref{uvt_plot} between the 1$\times$ and 
30$\times$ solar cases.  The overriding feature in all simulations is the development of a prograde (eastward) flow at 
low pressure.  The flow patterns in the 30$\times$ solar case exhibit clear wavelike structures outside of the equatorial 
region at the 1 bar, 30 mbar, and 1 mbar levels.  At the 1 bar level, the flow is predominately westward in the 30$\times$ solar 
case and predominately eastward in the 1$\times$ solar case.  In both the 1$\times$ and 30$\times$ solar metallicity cases the 
strength of the winds indicated by the length of the wind vectors in Figure~\ref{uvt_plot} is a stronger function of latitude than 
longitude, except at the highest levels (1 mbar) in the 30$\times$ solar case, which shows a significant day/night temperature 
contrast similar to what was seen in the simulations of the tidally locked hot Jupiters HD189733b and HD209458b from \citet{sho09}.  

We find that the atmospheric metallicity plays a key role in determining the jet structure for a planet 
with temperatures similar to those expected on GJ436b.  This is demonstrated clearly in Figure~\ref{uave}, which
shows the zonal-mean zonal wind\footnote{That is, the longitudinally averaged east-west wind, where eastward is defined
positive and westward negative.  See \citet{hol04}.} versus latitude and pressure for each of the five atmospheric 
metallicities for GJ436b considered in this study.  In the 1$\times$ solar case, strong high-latitude jets develop in the 
atmosphere with a weaker equatorial jet.  Increasing the metallicity of the planet to 3$\times$ solar causes a strengthening of the equatorial 
jet and a weakening of the high-latitude jets.  Once the metallicity of the atmosphere is increased to 10$\times$ solar or more, the high-latitude jets 
disappear and the equatorial jet becomes dominant.  Overall, the maximum zonal wind speed increases with metallicity from 
roughly 1300 m s$^{-1}$ in the 1$\times$ solar case to over 2000 m s$^{-1}$ in the 50$\times$ solar case.  
The flow is subsonic everywhere in our 1, 3, and 10$\times$ solar metallicity cases.  In our 30 and 50$\times$ solar metallicity cases,
the winds are subsonic throughout most of the domain, but they become marginally supersonic at the very top of the domain 
(at pressures $\sim$1 mbar) in the equatorial region.  Hydraulic jumps similar to those seen in the HD189733b and HD209458b 
cases presented in \citet{sho09} are present in the supersonic regions of the atmosphere in the 30 and 50$\times$ solar 
metallicity cases (see 1 mbar level of the 30$\times$ solar case in Figure \ref{uvt_plot}). It is important to note that the 
flow in all of the metallicity cases are predominately eastward at pressure less than $\sim$1 bar and predominately westward 
at pressures greater than $\sim$1 bar.  Momentum conservation requires that the eastward momentum in the upper atmosphere of 
these simulations comes from the deeper atmospheric layer, which requires the development of the mean westward flow at depth.  
The detailed mechanisms responsible for this momentum and the jet pumping mechanisms themselves will be discussed in a future paper.

In rapidly rotating atmospheres, atmospheric temperature gradients are linked to winds by dynamical balances, so it is interesting to next examine
the atmospheric temperature structure in our simulations.  Figure~\ref{tave} shows the zonal-mean atmospheric temperature versus latitude and pressure for
the 1$\times$ and 30$\times$ solar cases.  In both cases, the deepest isotherms (for temperatures exceeding $\sim$1200 K) are flat, but isotherms 
between 600 and 1100 K are bowed upward, indicating a warm equator and cool poles.  The relationship between this structure and the winds can be understood
with the thermal-wind equation, which relates the latitudinal temperature gradients to the zonal wind and its derivative with pressure:
\begin{equation}
\left(\frac{2u\tan\phi}{a}+2\Omega\sin\phi\right)\frac{\partial u}{\partial \ln p}=\frac{R}{a}\frac{\partial T}{\partial \phi},\label{thermwind}
\end{equation}
where $u$, $\phi$, $a$, $\Omega$, $p$, $R$ and $T$ are the zonal wind speed, latitude, planetary radius, planetary rotation rate, pressure, 
specific gas constant, and temperature, respectively.  The latitudinal temperature gradient on the right-hand side is evaluated at constant 
pressure.  This relationship derives from taking a vertical (pressure) derivative of the meridional momentum equation for a flow
where the predominant zonal-mean meridional momentum balance is between the Coriolis, pressure-gradient, and curvature terms 
(called ``gradient-wind'' balance; see Holton 2004, p. 65-68).  From Equation (\ref{thermwind}), one expects that, away from
the equator, regions exhibiting vertical shear of the zonal wind must also exhibit latitudinal gradients of temperature.  
Comparing Figures~\ref{uave} and \ref{tave} confirms that this is indeed the case: in the mid-latitudes, the 1$\times$ solar 
case exhibits the greatest vertical shear of the zonal wind in the pressure range of $\sim$0.1 to 3 bars (Figure~\ref{uave}),
and this is the same pressure range over which the greatest latitudinal temperature gradients occur (Figure~\ref{tave}).   
For the 30$\times$ solar case, in the mid-latitudes, the regions exhibiting significant wind shear are shifted upward, 
occurring from $\sim$0.01 to less than 1 bar, and likewise this is the pressure range where significant latitudinal 
temperature gradients exist.  The upward shift in the temperature gradients (and hence winds) at greater metallicity is the 
direct result of enhanced atmospheric opacities, which lead to shallower atmospheric heating \citep{for08, dob08}.

It is interesting to characterize the variations in wind speed that occur throughout the eccentric orbit due to the time-variable
incident stellar flux.  Figure~\ref{vrms_plot_hc} shows the RMS wind speed as a function of pressure and simulated time with outputs every 
two hours for a ten Earth day period after the simulations had reached equilibrium.  Overall, wind speeds in these simulations of GJ436b vary 
with a frequency roughly equal to the orbital period at pressures above the mean photospheric level\footnote{In
Figure \ref{vrms_plot}, the high-frequency fluctuations appear to occur on periods of tens of Earth days,
but this is an artifact that results from aliasing of the sampling frequency of $5\times10^5$ s with the orbital period.}
(roughly 100 mbar in the 1$\times$ solar case and 10 mbar in the 50$\times$ solar case).  Wind speeds
are fairly constant at pressures below the mean photosphere for each metallicity case of GJ436b.  
The vertical lines in Figure~\ref{vrms_plot_hc} indicate the time of periapse passage, which are followed by a peak in the RMS wind speeds.  
The variation in the wind speeds between periapse and apoapse is several times larger in the 50$\times$ solar 
case compared with the 1$\times$ solar case.  The higher atmospheric metallicity cases show greater variability in wind speeds as function of 
orbital phase, which could affect light curves especially if hotter or more eccentric systems are considered.

\subsection{Effect of eccentricity and rotation rate}

The models presented in Section~\ref{metallicity} assume an eccentric orbit (hence time-variable
stellar irradiation) and adopt the pseudo-synchronous rotation rate given in Table~\ref{gj436_params}.
Here, we explore the effect of eccentricity and rotation rate on the circulation by comparing our
standard cases (Section~\ref{metallicity}) to cases with synchronous (rather
than pseudo-synchronous) rotation rates and zero eccentricity.
Figure~\ref{uvt_plot_synch} presents the thermal structure and winds at the 30 mbar level for the 1$\times$ and 
30$\times$ solar cases assuming synchronous rotation ($P_{rot}=P_{orb}$).  The top panels of Figure~\ref{uvt_plot_synch} 
assume the nominal eccentric orbit of GJ436b while the bottom panels assume a circular orbit with the nominal semimajor 
axis of GJ436b (Table~\ref{gj436_params}).  The assumption of synchronous rotation and/or a circular orbit has little 
effect on the overall thermal structure and wind patterns that develop in these simulations.  This is presumably because 
the eccentricity of GJ436b's orbit is modest ($e=0.15$) and changes in the average stellar flux and pseudo-synchronous 
rotation rate from a circularized and tidally locked orbit are small.  

Figure~\ref{uave_synch} presents the zonal-mean zonal winds for these same four cases (1$\times$ and 
30$\times$ solar metallicity, with eccentricity of 0 or 0.15, all using the synchronous rotation period).
Overall, the jet structures differ little from the nominal cases.  
It is interesting to note that the eastward equatorial jet in the 30$\times$ solar synchronous rotation 
cases has a maximum wind speed that is $\sim 200$ m s$^{-1}$ faster than what is seen in the non-synchronous case.  Because 
GJ436b has a relatively small eccentricity orbit the effects of non-synchronous rotation and time-variable heating are only 
small perturbations on the synchronous rotation and circular orbit cases.   Planets with higher eccentricities are likely to show 
a larger variation in the circulation patterns that develop compared with circularized and synchronous cases.
 
\subsection{Light Curves and Spectra}\label{spec}

\begin{figure}
\centering
\includegraphics[width=0.49\textwidth]{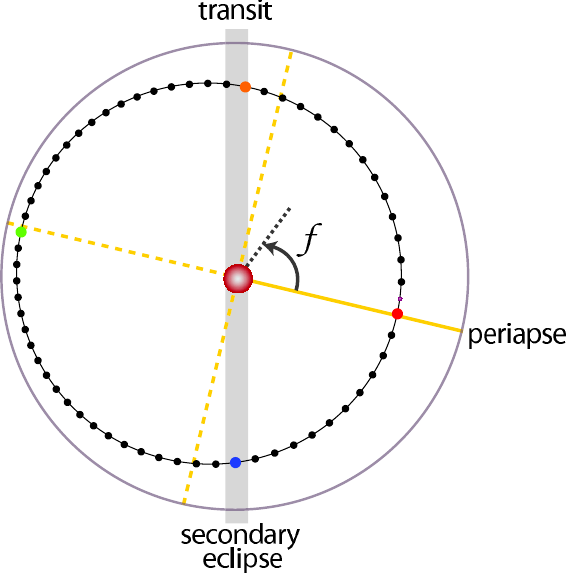}
\caption{Orbit of GJ436b.  The true anomaly, $f$, represents the angular distance of the planet from periapse. Assuming a longitude 
of pericenter, $\varpi$, of 343$^{\circ}$ from \citet{demi07}, transit occurs at $f=107^{\circ}$ and secondary eclipse occurs at   
$f=-73^{\circ}$.  Dots along the orbital path represent points where data was extracted to produce Figures \ref{light_curves}, 
\ref{fluxes}, and \ref{energy}.  Colored dots represent points near periapse (red), transit (orange), apoapse (green), and 
secondary eclipse (blue), which correspond to the colored spectra presented in Figures \ref{fluxes} and \ref{energy}.  Figure 
is to scale with the small purple dot after periapse representing the size of GJ436b in relation to its host star and orbit.
(a color version of this figure is available in the online journal.)}\label{orbit_fig}
\end{figure}

\begin{figure}
\centering
  \includegraphics[width=0.49\textwidth]{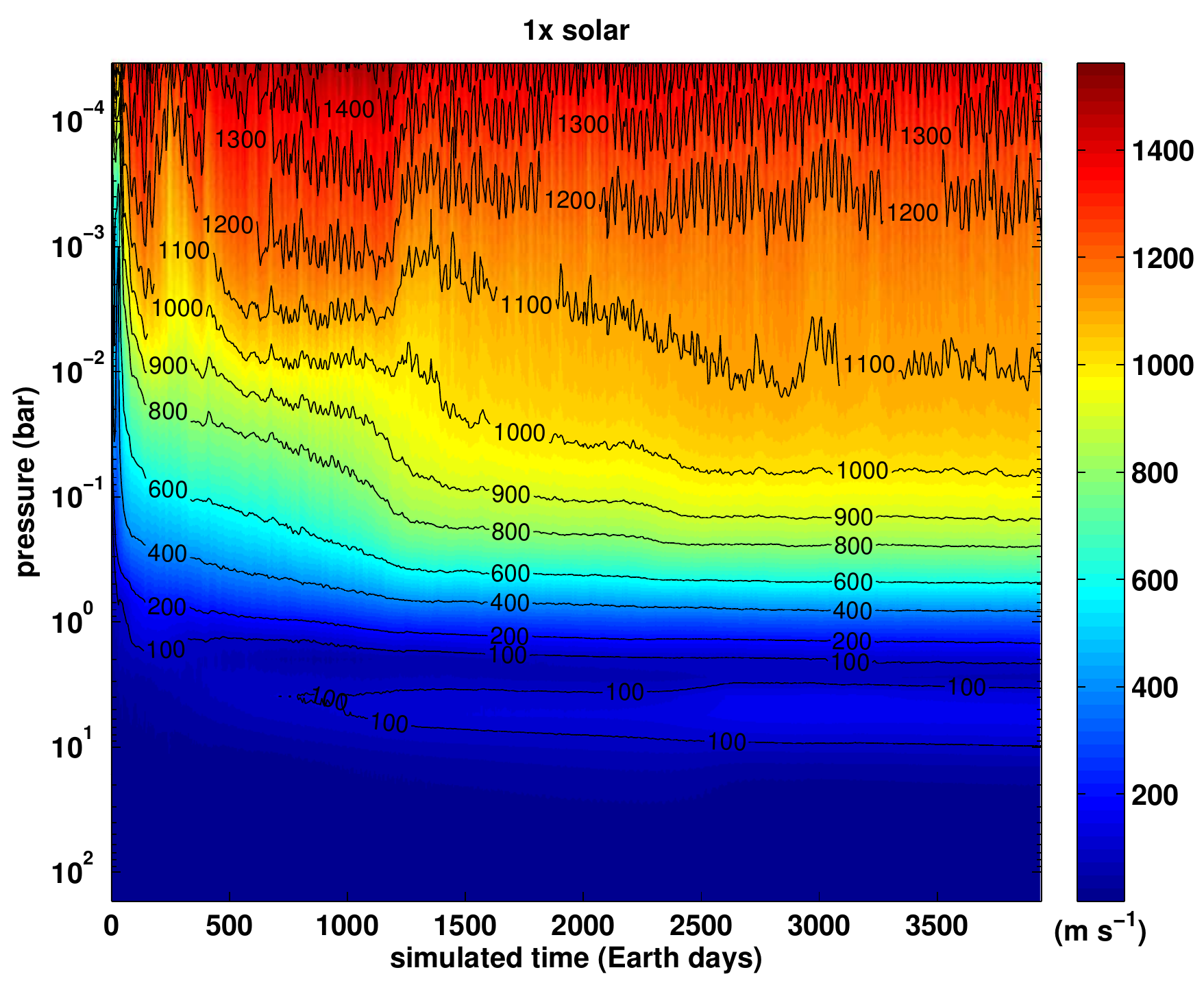}\\
  \includegraphics[width=0.49\textwidth]{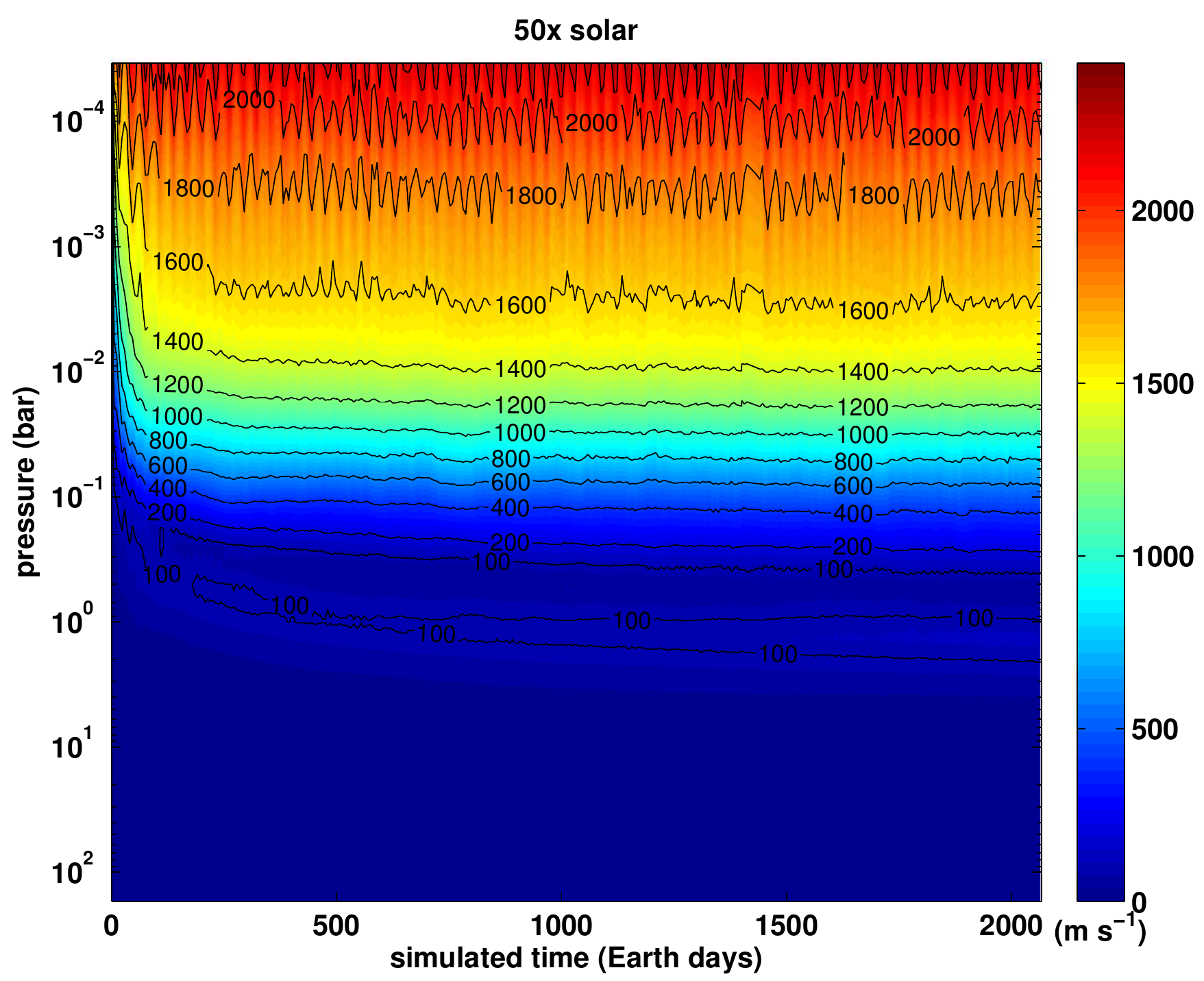}
  \caption{RMS velocity (colorscale) as a function of pressure and simulated time for the 1$\times$ (top) and 50$\times$ (bottom) solar cases of GJ436b. 
           The RMS velocity at each pressure level is calculated from the instantaneous wind speeds recorded every 5$\times$10$^5$ s.  Simulations 
           are continued until the winds reach a relatively flat profile at all pressure levels. 
           (a color version of this figure is available in the online journal.)}\label{vrms_plot}
\end{figure}

\begin{figure}
 \centering
   \includegraphics[width=0.45\textwidth]{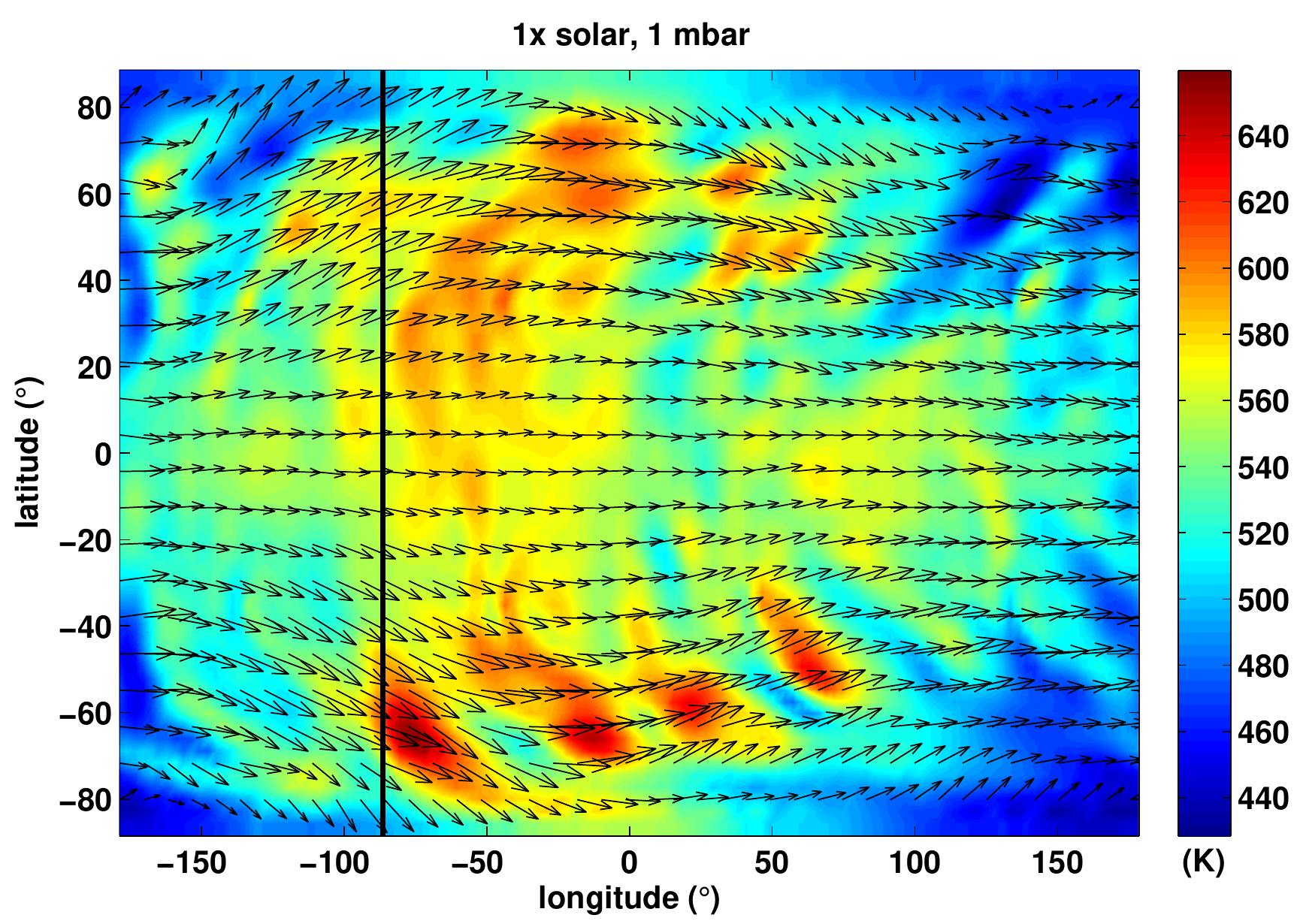}
   \includegraphics[width=0.45\textwidth]{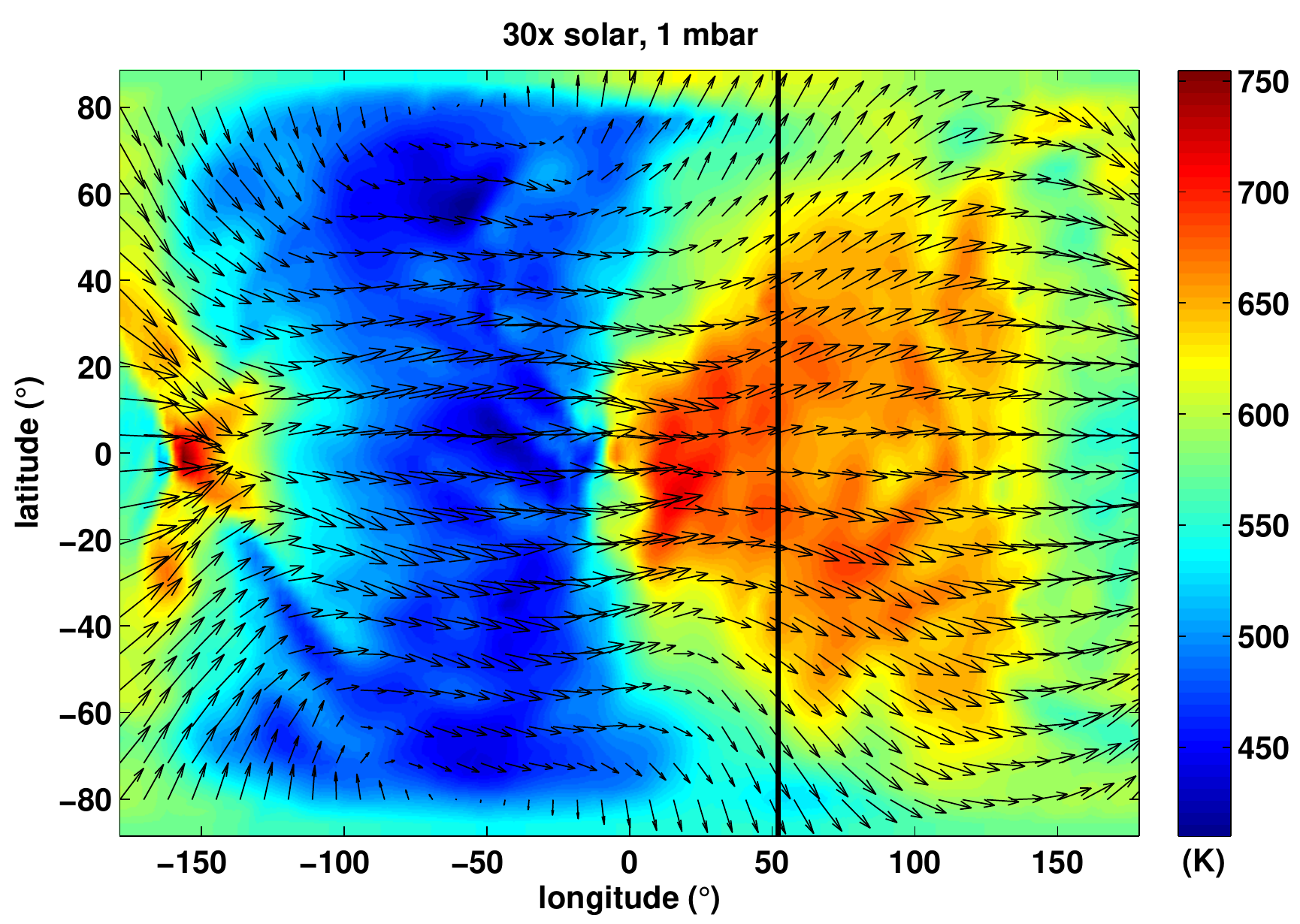}\\
   \includegraphics[width=0.45\textwidth]{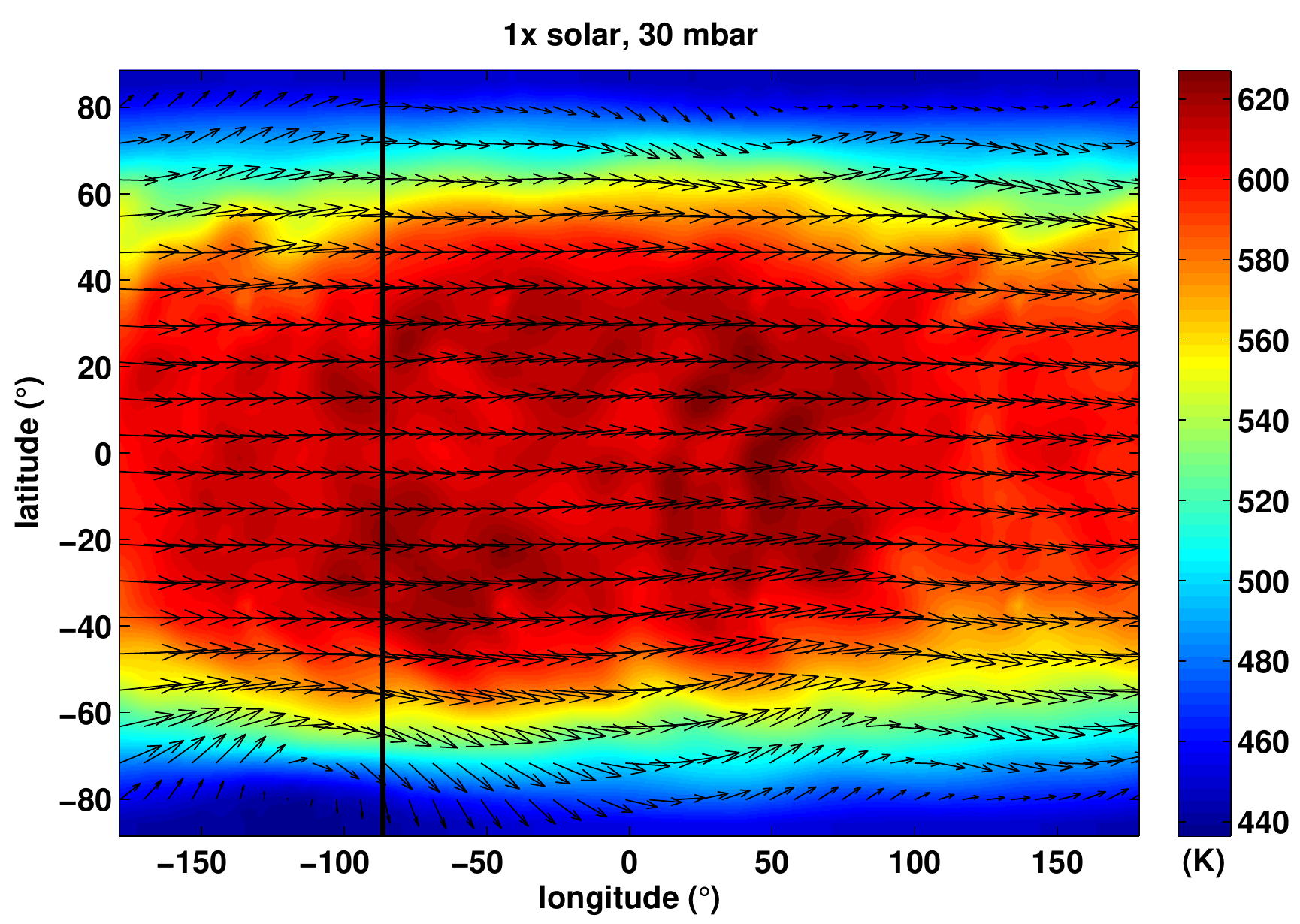}
   \includegraphics[width=0.45\textwidth]{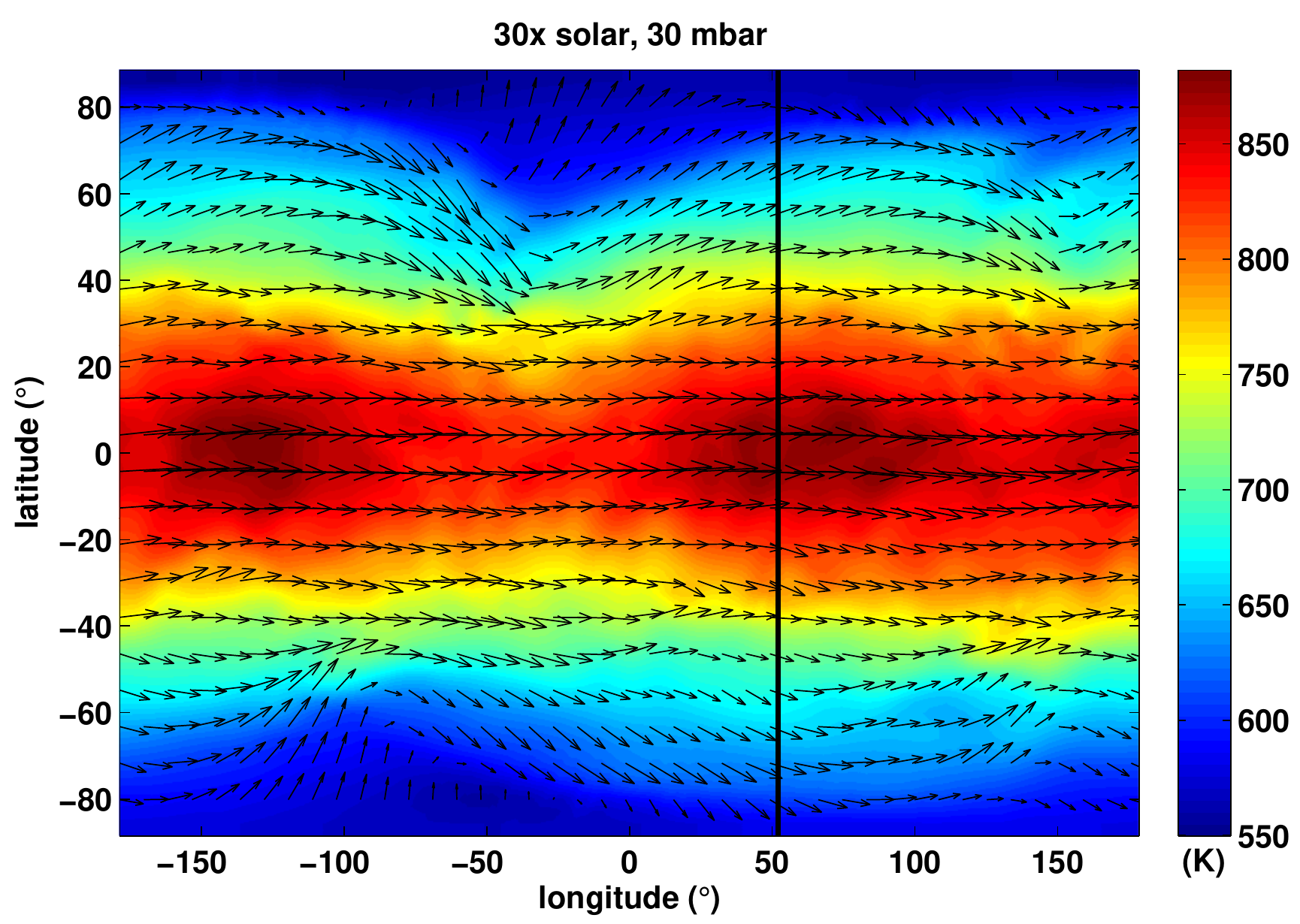}\\
   \includegraphics[width=0.45\textwidth]{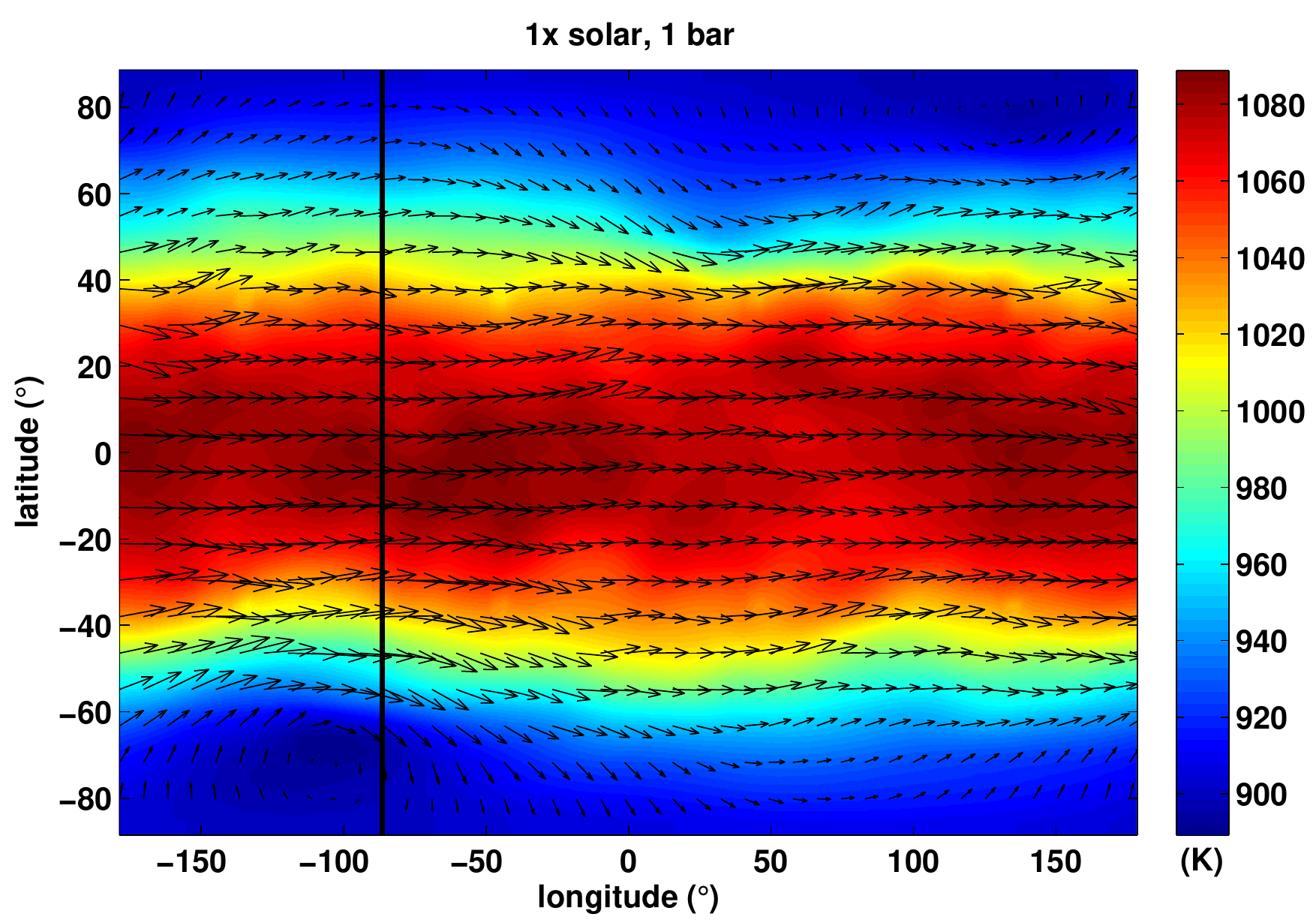}
   \includegraphics[width=0.45\textwidth]{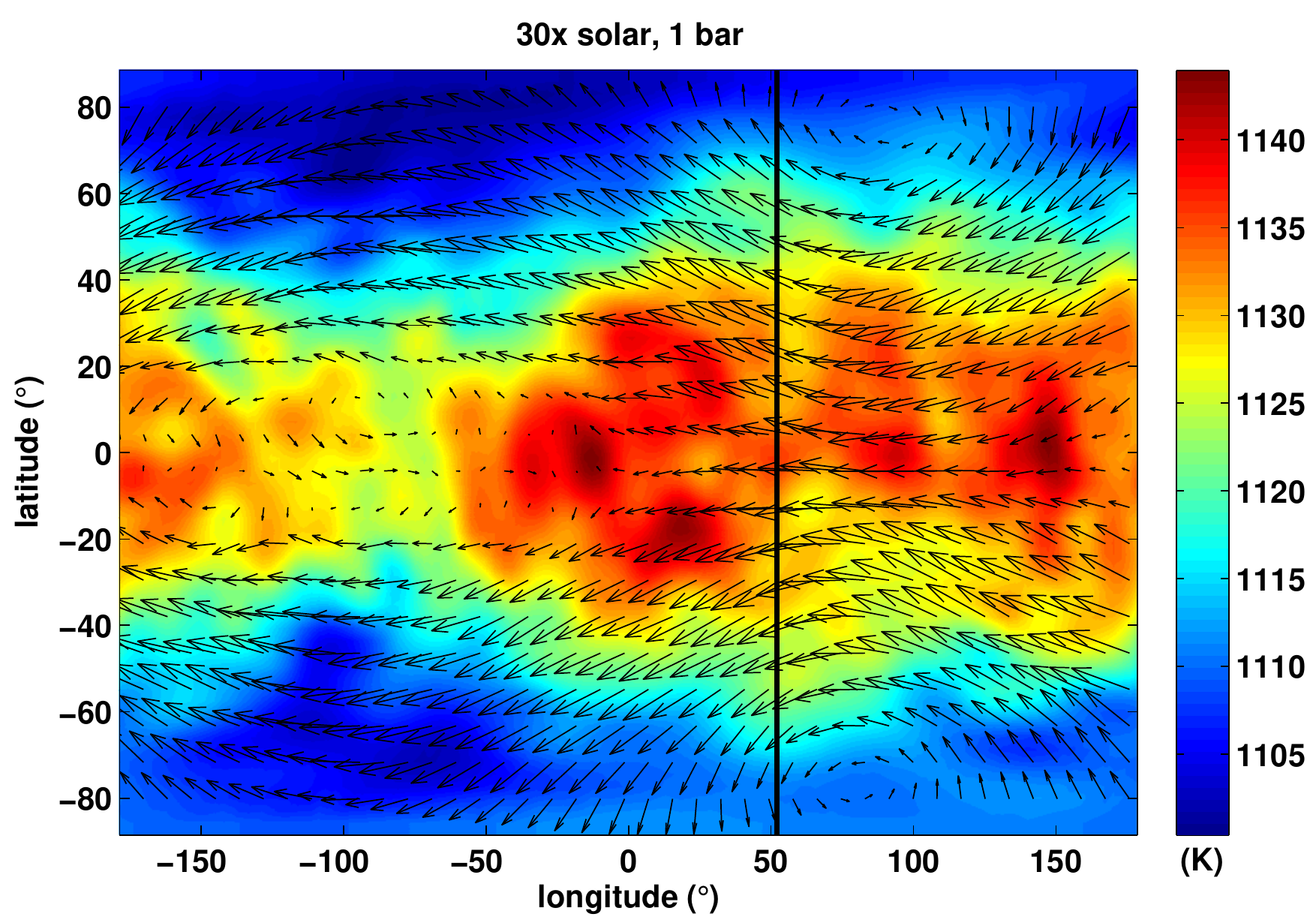}
\caption{Temperature (colorscale) and winds (arrows) for the 1$\times$ (left) and 30$\times$ (right) solar metallicity cases of GJ436b.  
For both 1$\times$ and 30$\times$ solar cases the thermal structure and winds are shown at the 1 mbar (top), 30 mbar (middle), 
and 1 bar (bottom) levels of the simulation.  The longitude of the substellar point is indicated by the solid vertical line 
in each panel.  Each panel is a snap shot of the atmospheric state taken near secondary eclipse ($f = -73\degr$, Figure \ref{orbit_fig}).  The 
horizontal resolution of these runs is C32 (roughly 128$\times$64 in longitude and latitude) with 47 vertical layers.
(a color version of this figure is available in the online journal.)}\label{uvt_plot}
\end{figure}

\begin{figure}
 \centering
   \includegraphics[width=0.43\textwidth]{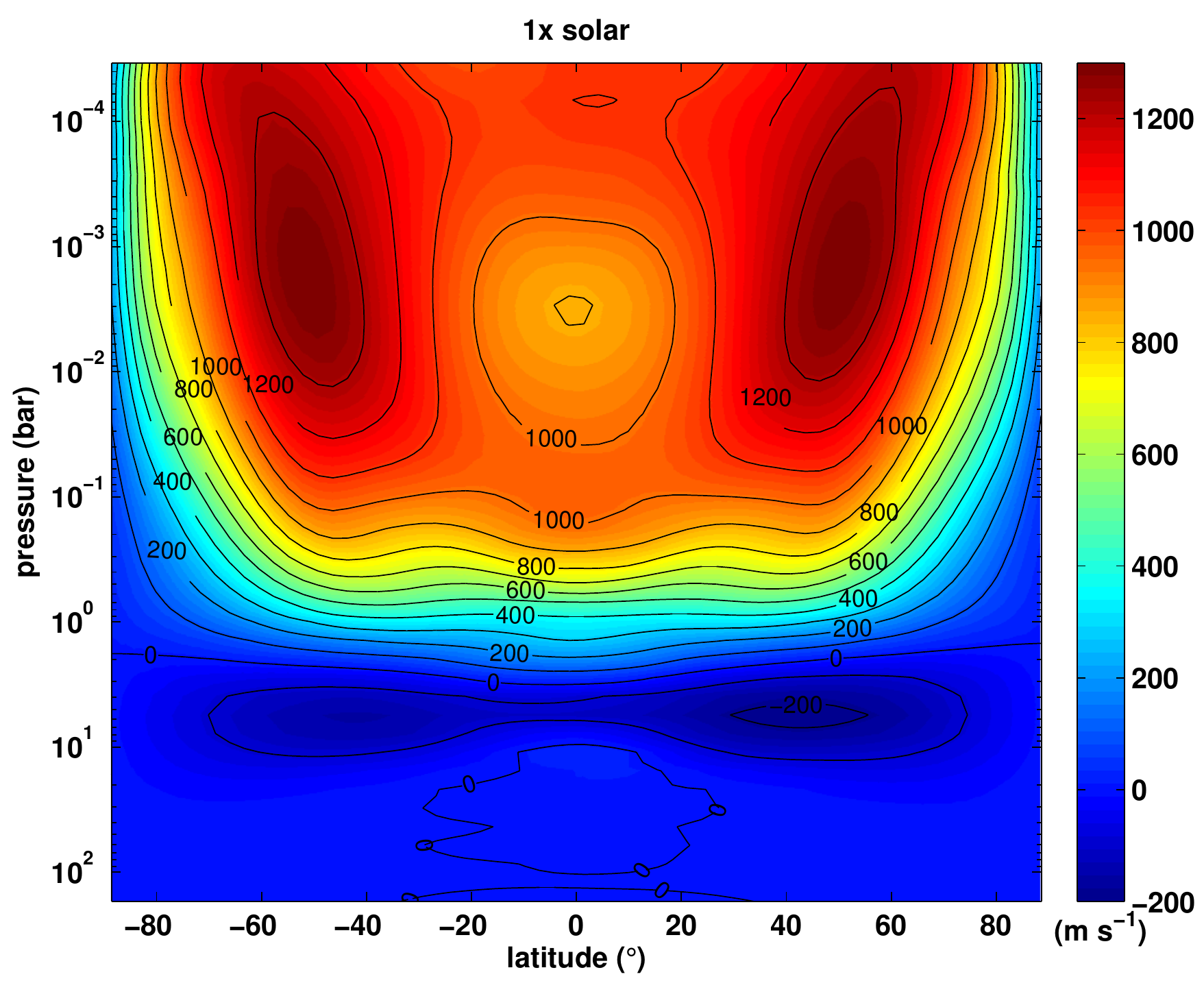}
   \includegraphics[width=0.43\textwidth]{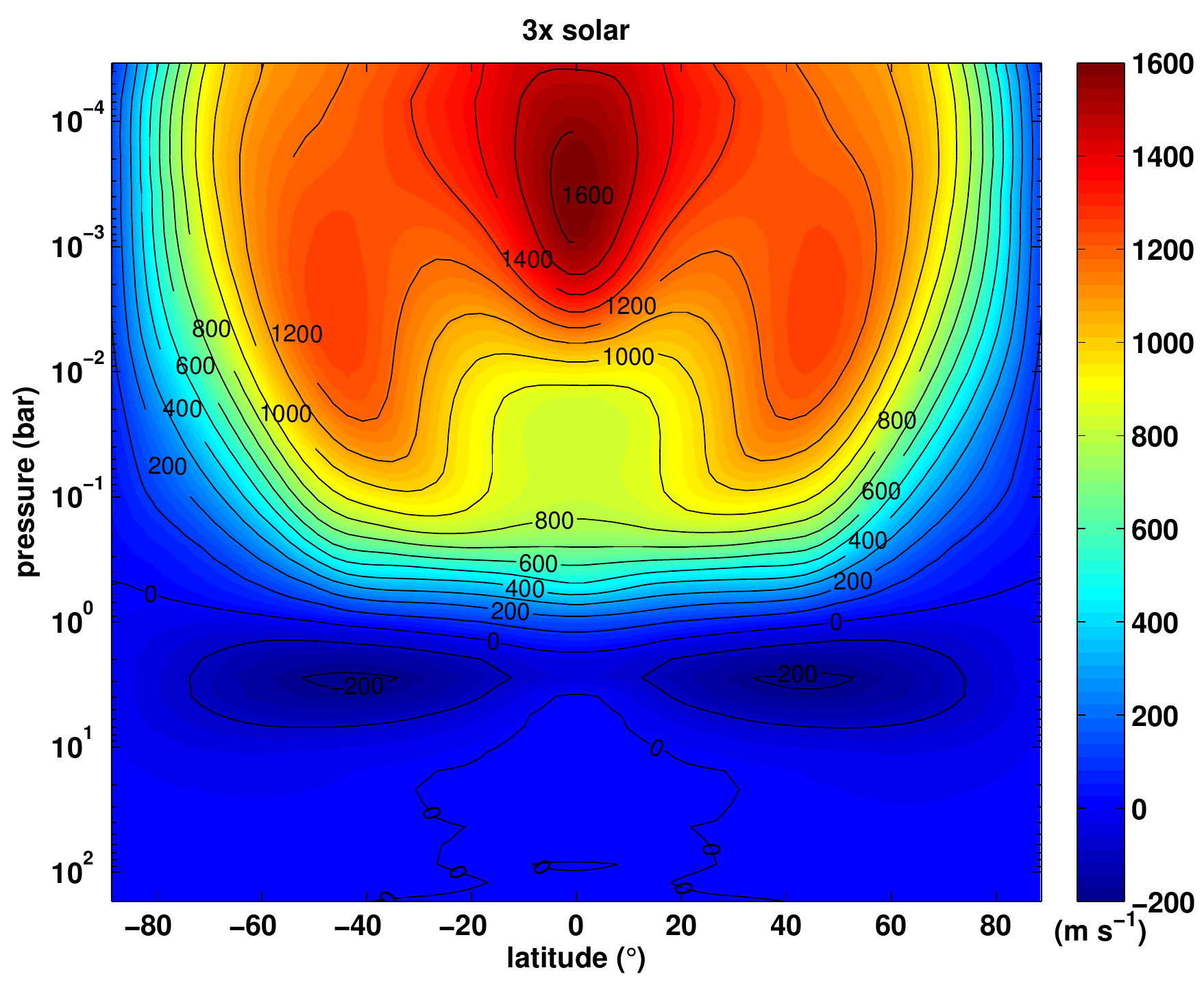}\\
   \includegraphics[width=0.43\textwidth]{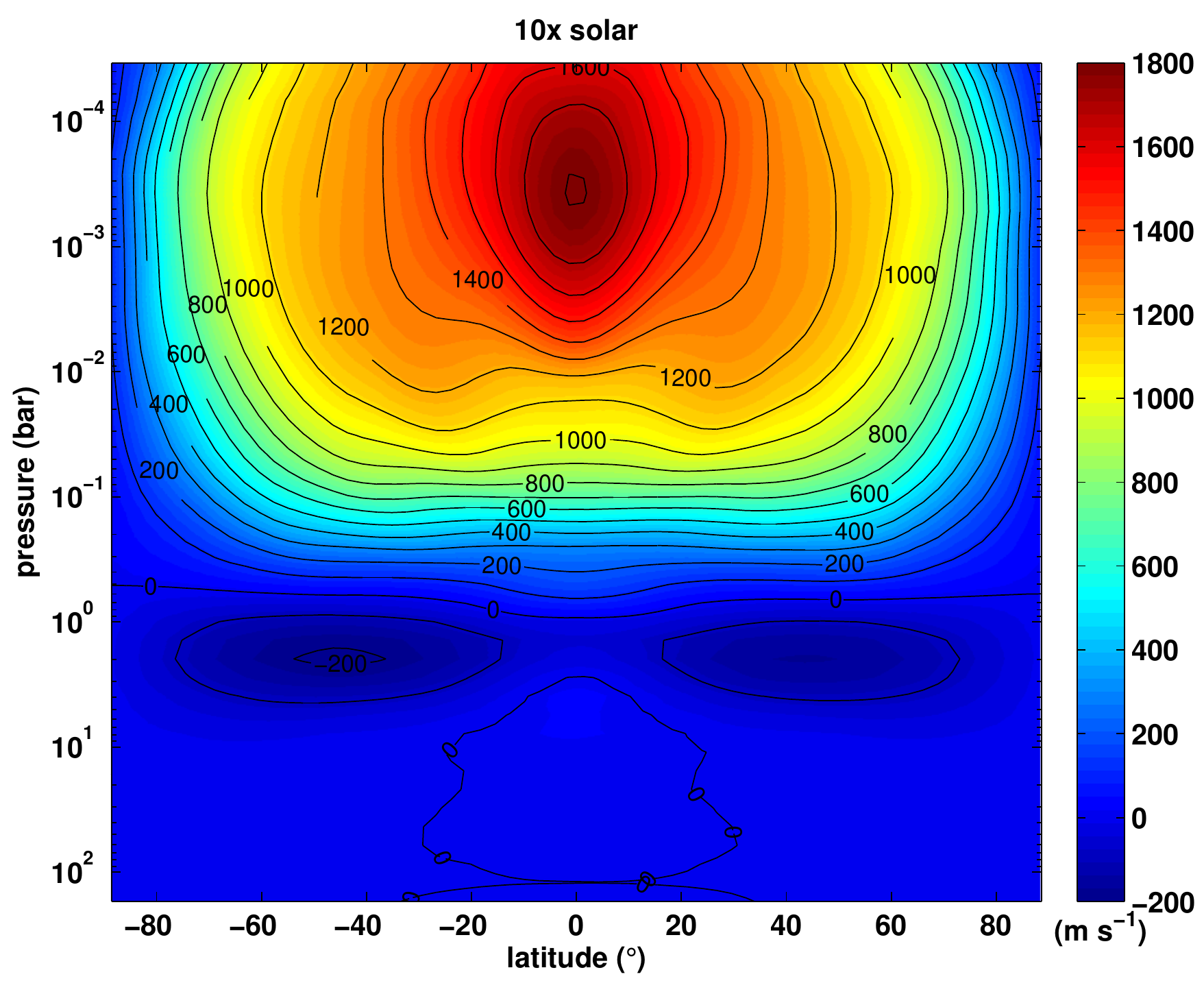}
   \includegraphics[width=0.43\textwidth]{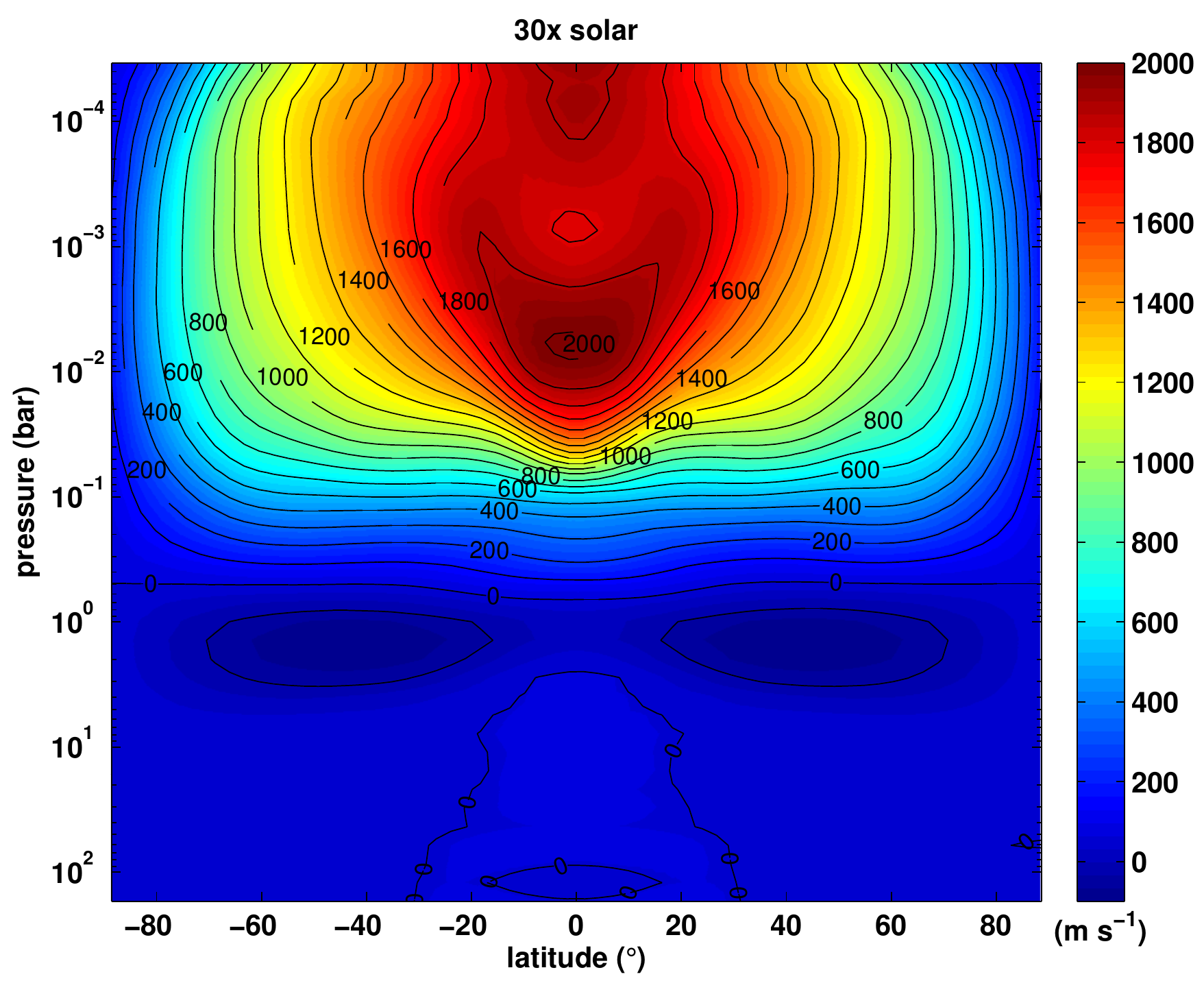}\\
   \includegraphics[width=0.43\textwidth]{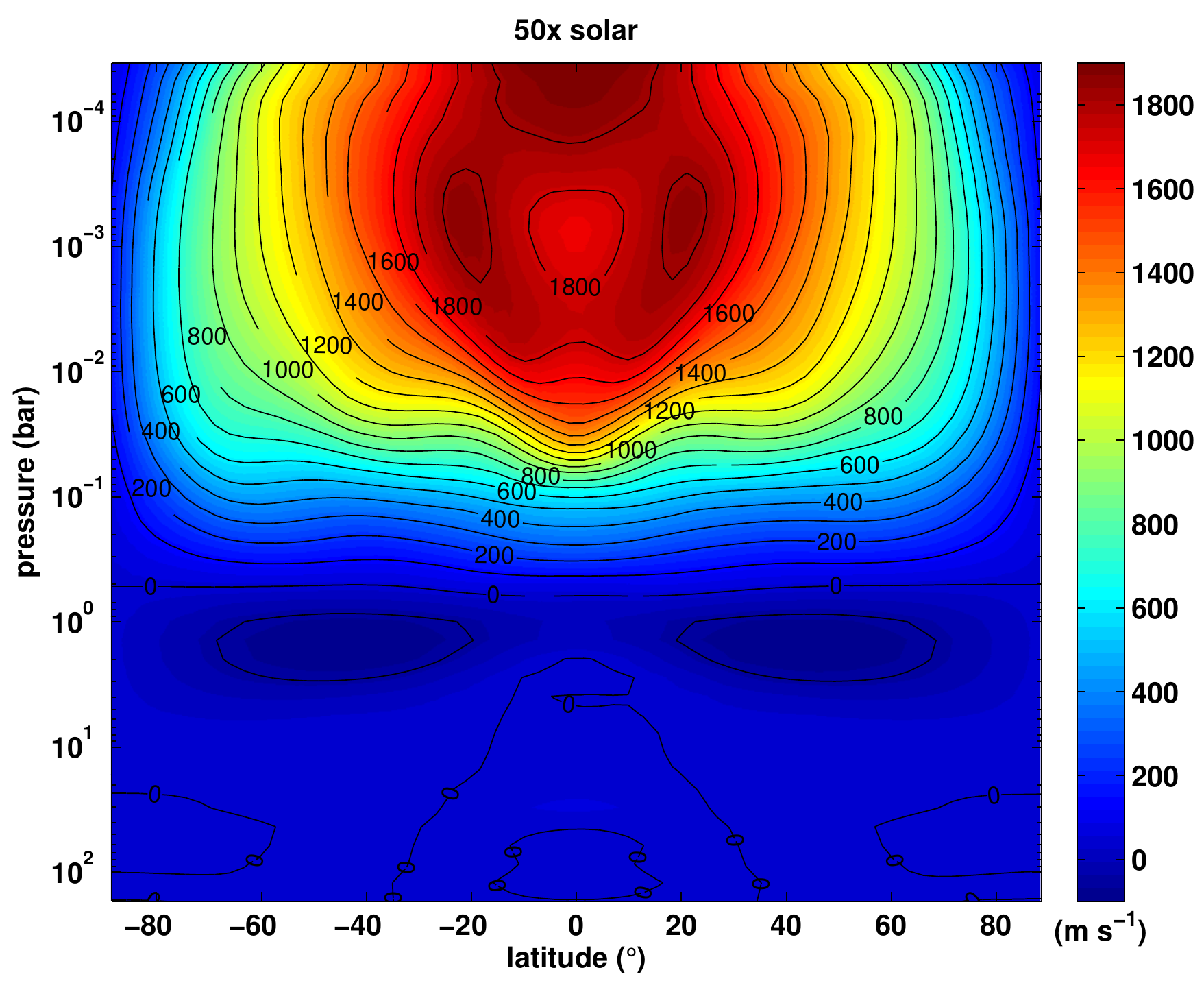}
\caption{Zonal-mean zonal winds for the five atmospheric metallicities  
considered in this study for GJ436b assuming pseudo-synchronous rotation.  The wind speeds presented here represent 100 day averages of the zonal 
winds taken after each simulation was considered to have reached an equilibrium state. The colorbar 
shows the strength of the zonally averaged winds in m s$^{-1}$.  Contours are spaced by 100 m s$^{-1}$.
Positive wind speeds are eastward, while negative wind speeds are westward.  Note the significant change in 
the jet structure as a function of atmospheric metallicity.(a color version of this figure is available in the online journal.)}\label{uave}
\end{figure}

\begin{figure}
\centering
  \includegraphics[width=0.49\textwidth]{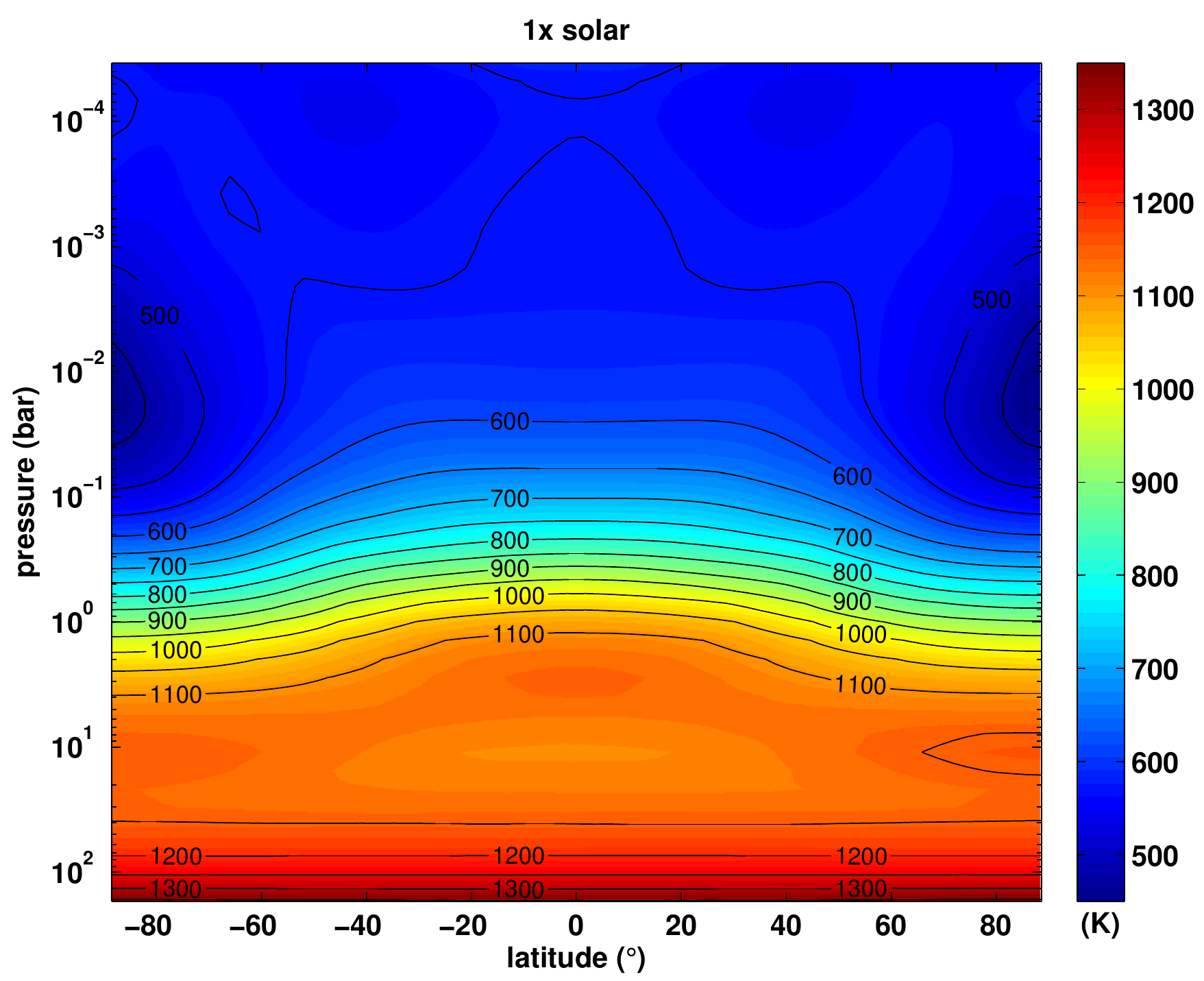}\\
  \includegraphics[width=0.49\textwidth]{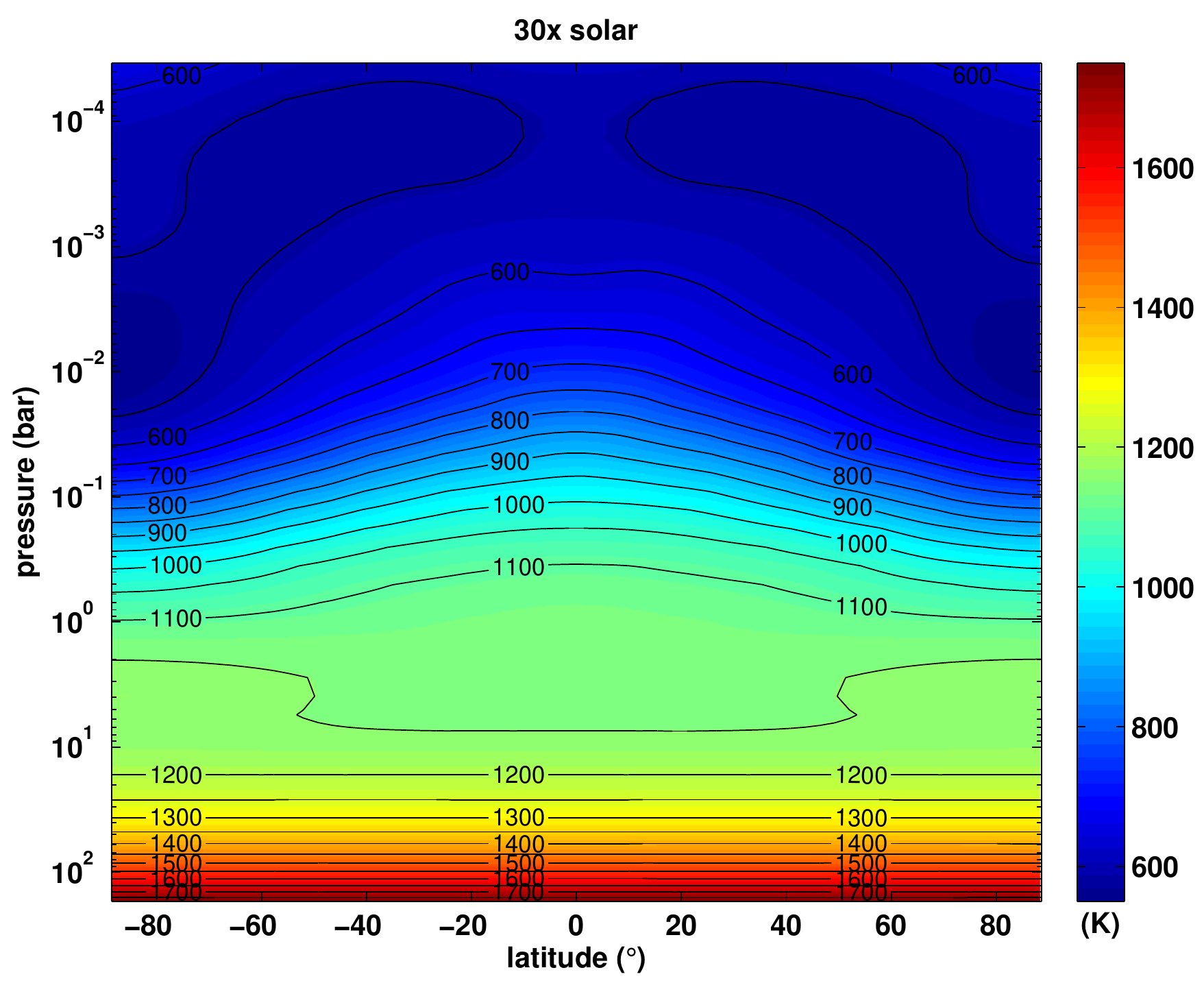}
  \caption{Zonal-mean temperatures (colorscale) as a function of pressure and latitude for the pseudo-synchronous 
           1$\times$ (top) and 30$\times$ (bottom) solar cases of GJ436b.  Contours represent isotherms and are spaced by 50 K.
           Gradients in temperature along an isobaric surface tend to drive zonal winds according to equation (\ref{thermwind}).  
           The 30$\times$ solar case shows stronger thermal gradients at lower pressures that extend into lower latitudes 
           when compared with the 1$\times$ solar case.  This change in the thermal gradient structure can be directly 
           related to the zonal wind profiles seen in Figure \ref{uave}.(a color version of this figure is available in the online journal.)}\label{tave}
\end{figure}

\begin{figure}
\centering
  \includegraphics[width=0.49\textwidth]{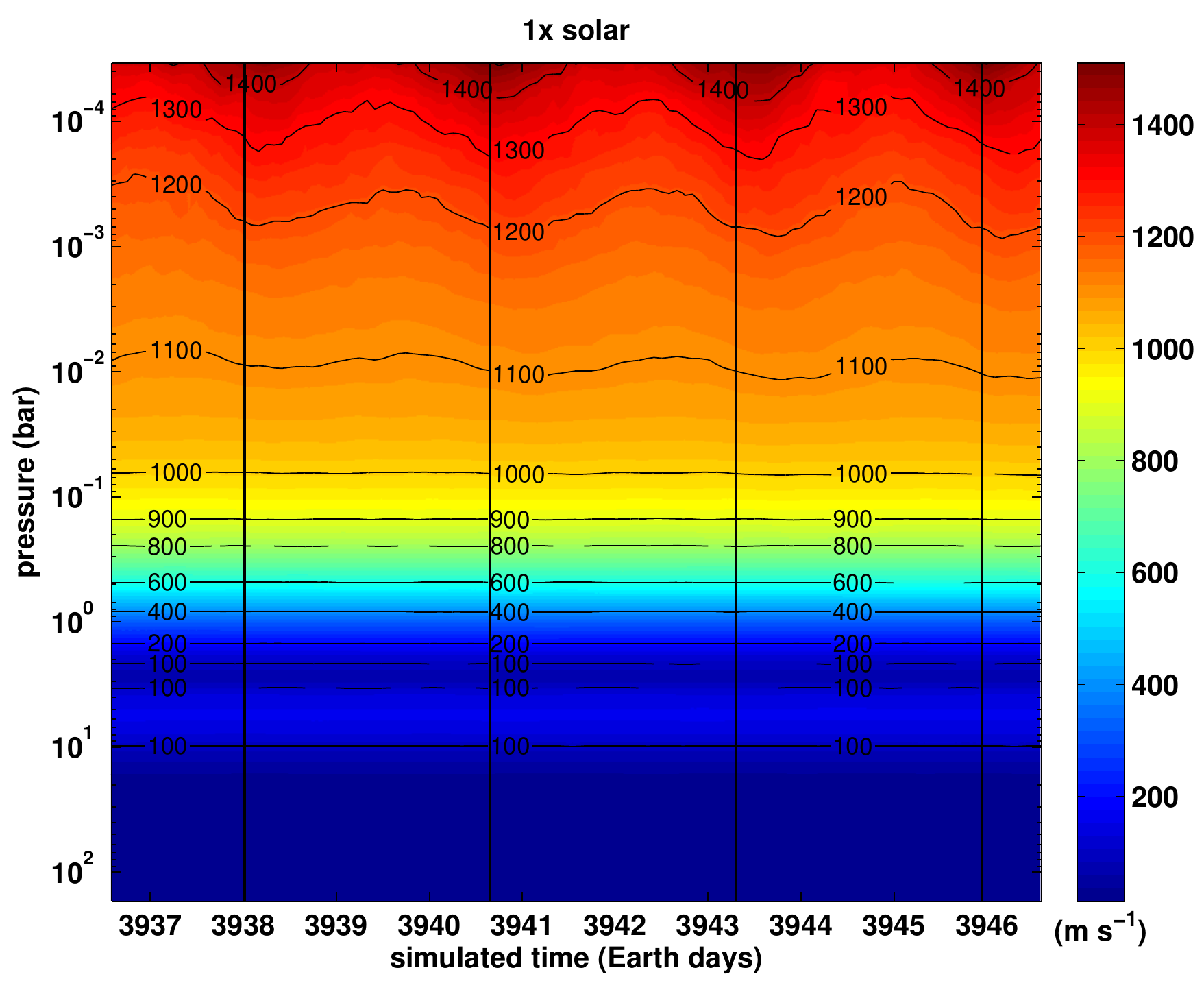}\\
  \includegraphics[width=0.49\textwidth]{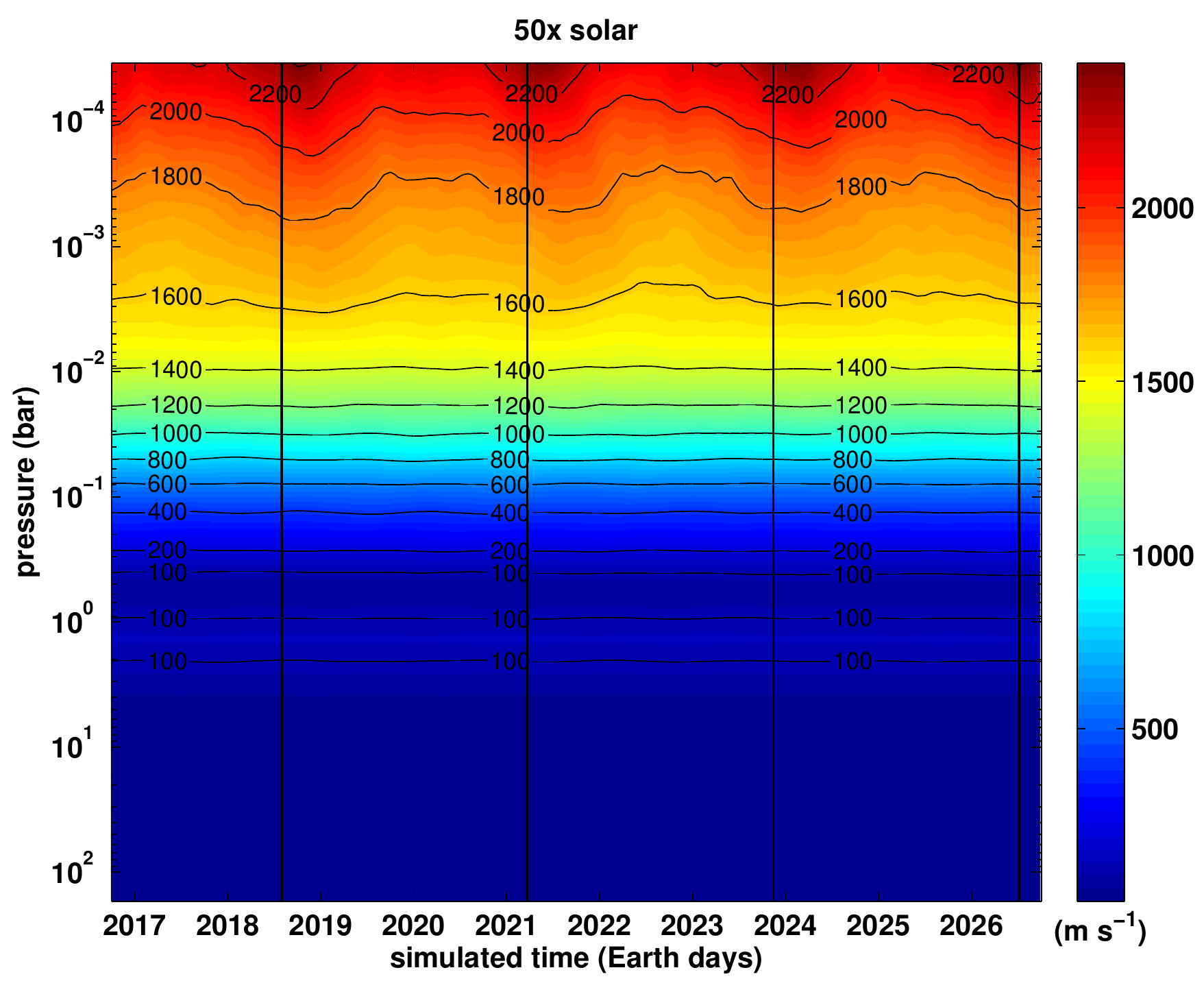}
  \caption{RMS velocity (colorscale) as a function of pressure and simulated time for the 1$\times$ (top) and 50$\times$ (bottom) 
           solar cases of GJ436b as calculated from high-cadence simulation outputs every 7200 s. Vertical 
           lines indicate periapse passage.  Overall, wind speeds in the atmosphere vary with a period 
           equal to the orbital period.  The peak in the wind speeds typically occurs 4-8 hours after periapse passage.
           (a color version of this figure is available in the online journal.)}\label{vrms_plot_hc}
\end{figure}

\begin{figure}
 \centering
   \includegraphics[width=0.45\textwidth]{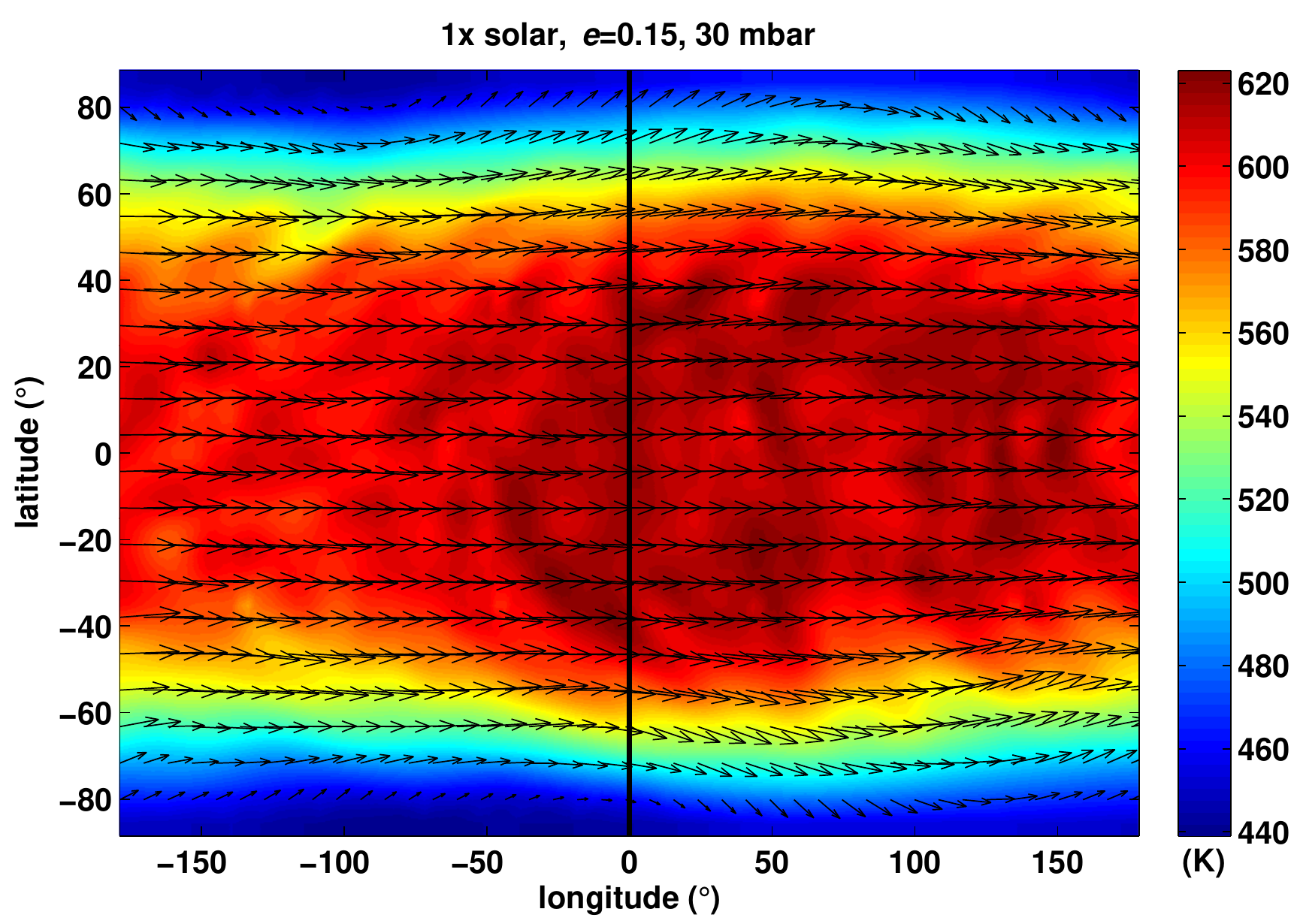}
   \includegraphics[width=0.45\textwidth]{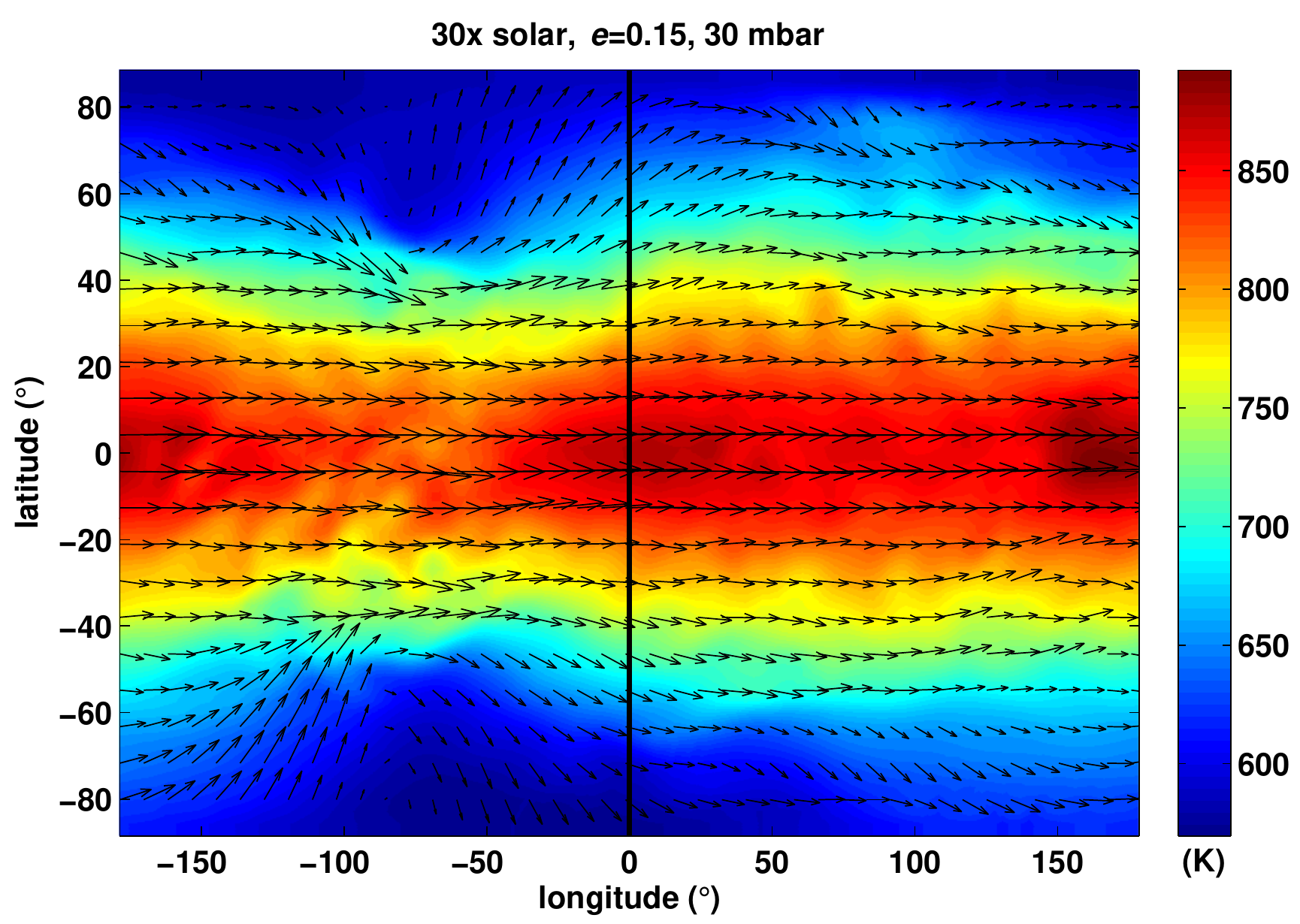}\\
   \includegraphics[width=0.45\textwidth]{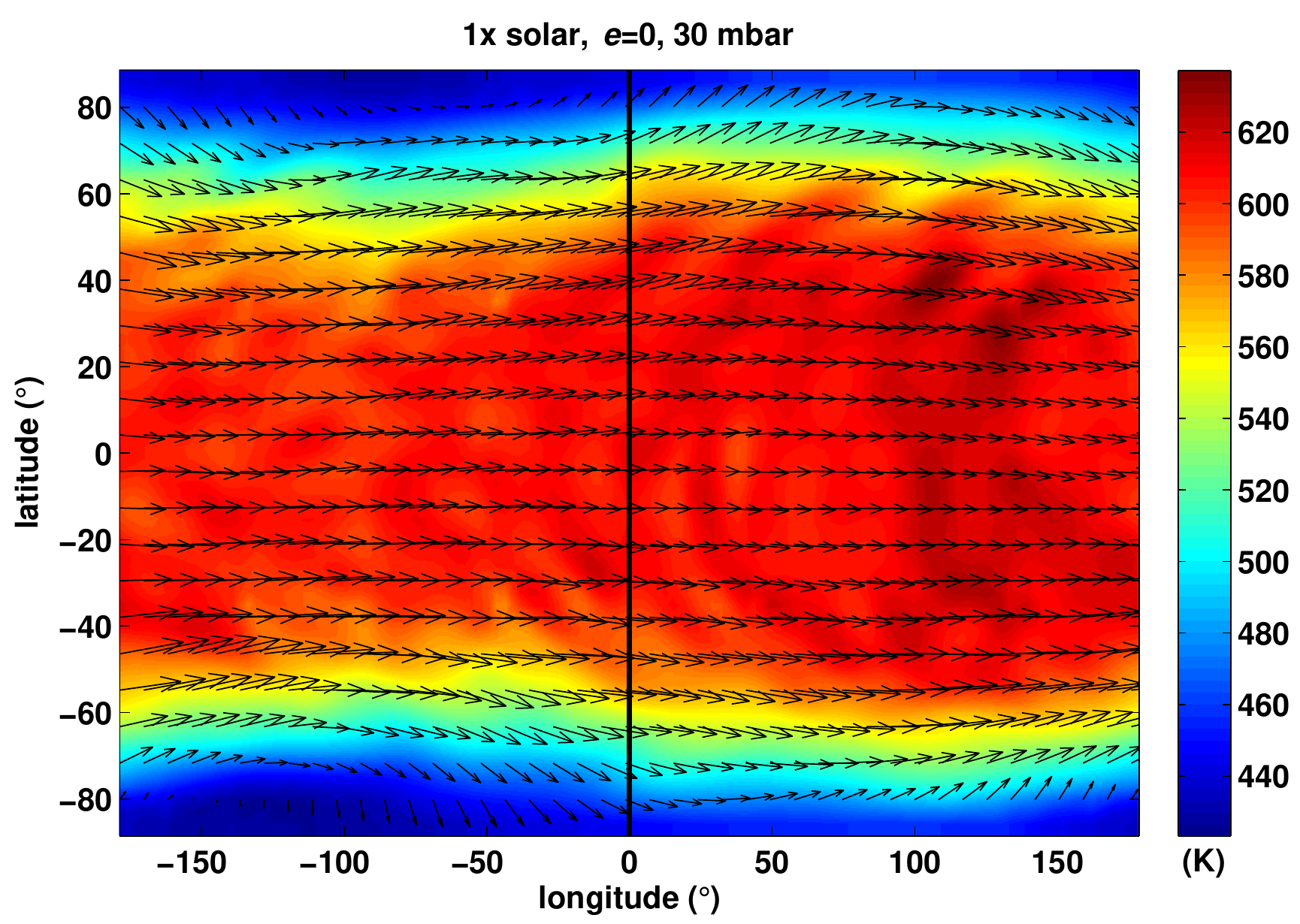}
   \includegraphics[width=0.45\textwidth]{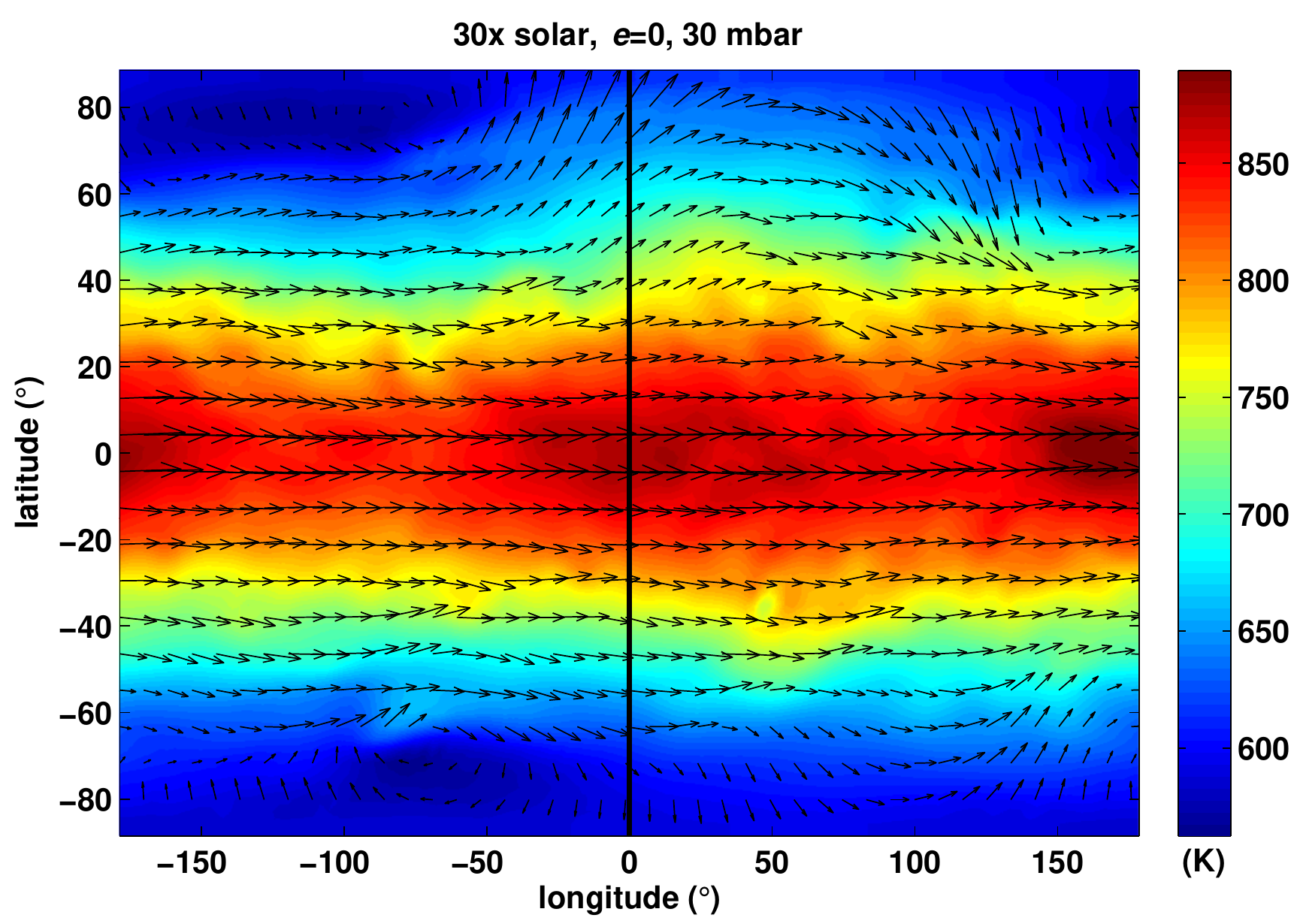}
\caption{Temperature (colorscale) and winds (arrows) for the 1$\times$ (left) and 30$\times$ (right) solar metallicity cases of GJ436b 
assuming synchronous rotation at the 30 mbar level.  The top panels represent simulations where synchronous rotation 
was assumed, but the nominal eccentric orbit was maintained.  The bottom panels represent simulations where synchronous 
rotation and a circular orbit with the nominal semimajor axis (Table \ref{gj436_params}) were assumed.  In the eccentric 
cases (top) the panels represent a snap shot taken near secondary eclipse ($f = -73\degr$, Figure \ref{orbit_fig}).  The solid vertical line 
in each panel represents the longitude of the substellar point.(a color version of this figure is available in the online journal.)}\label{uvt_plot_synch} 
\end{figure}

\begin{figure}
 \centering
   \includegraphics[width=0.45\textwidth]{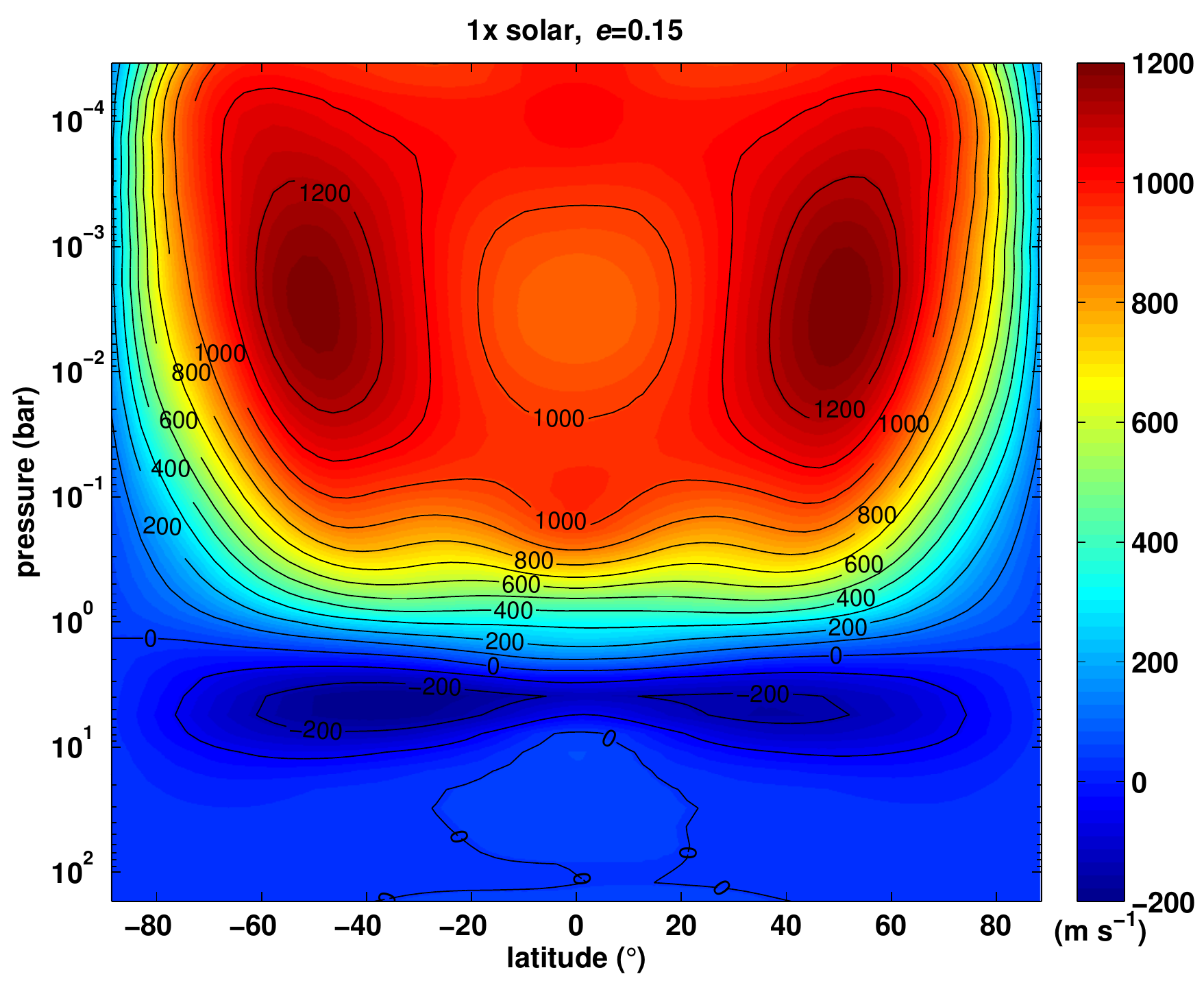}
   \includegraphics[width=0.45\textwidth]{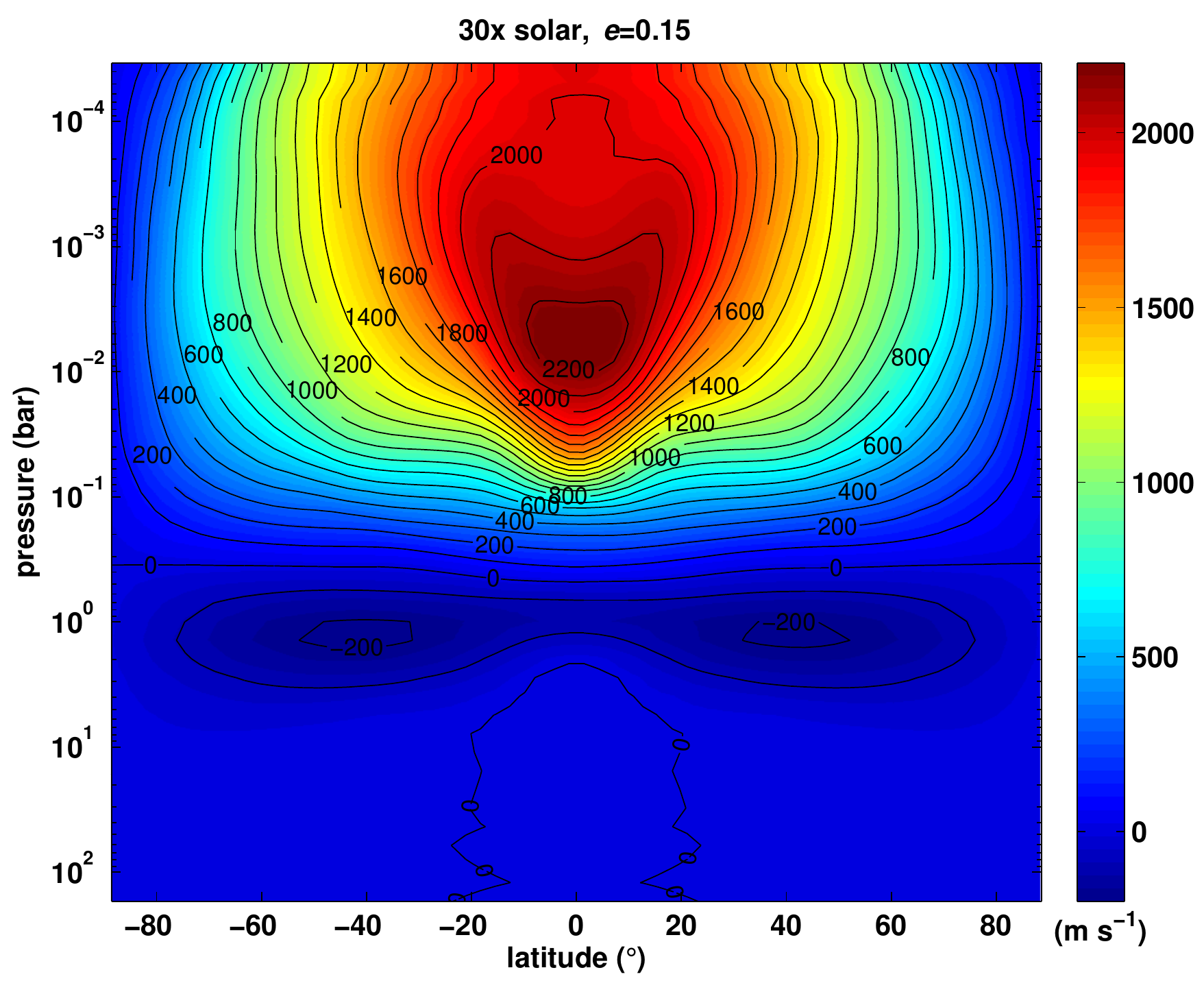}\\
   \includegraphics[width=0.45\textwidth]{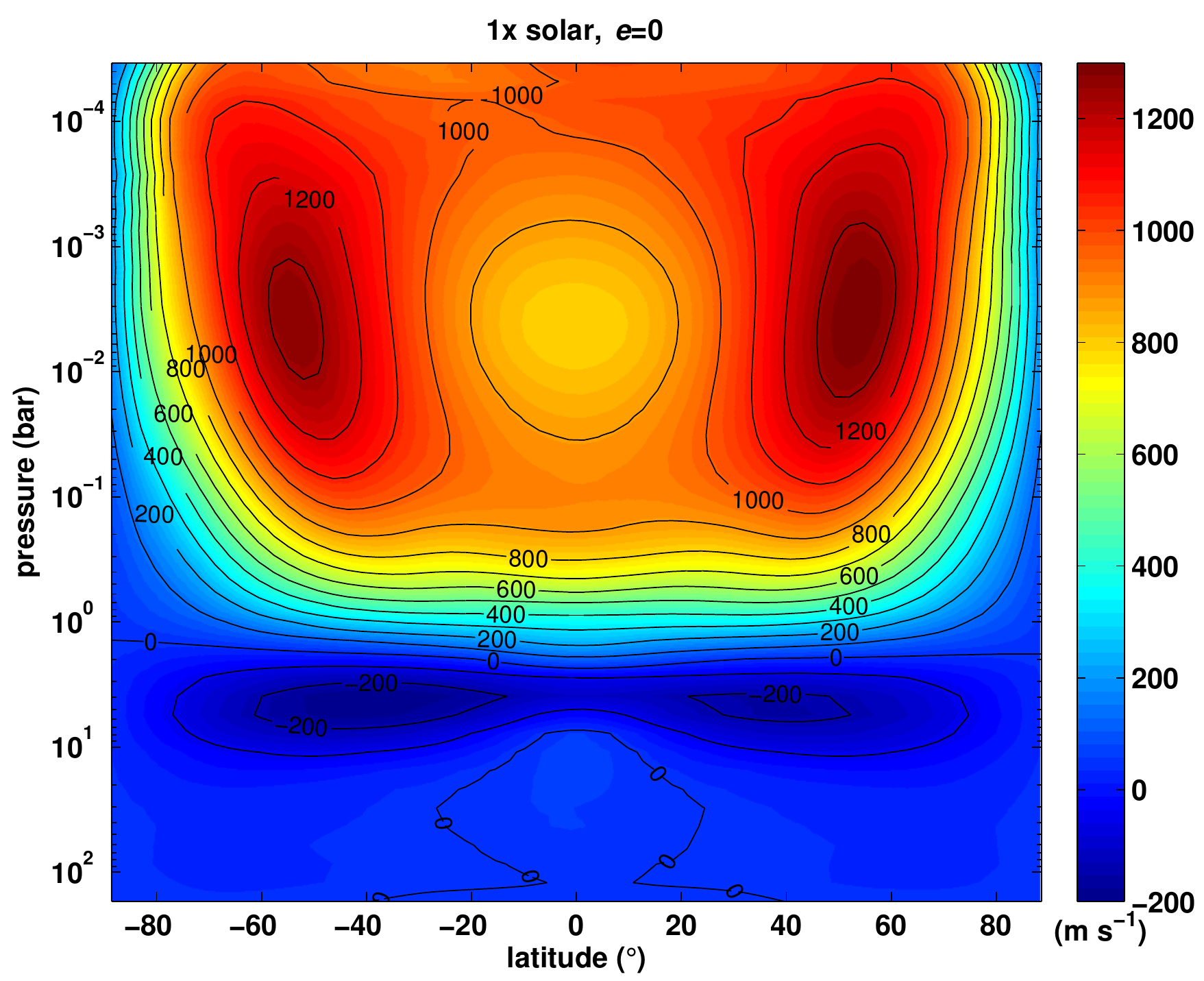}
   \includegraphics[width=0.45\textwidth]{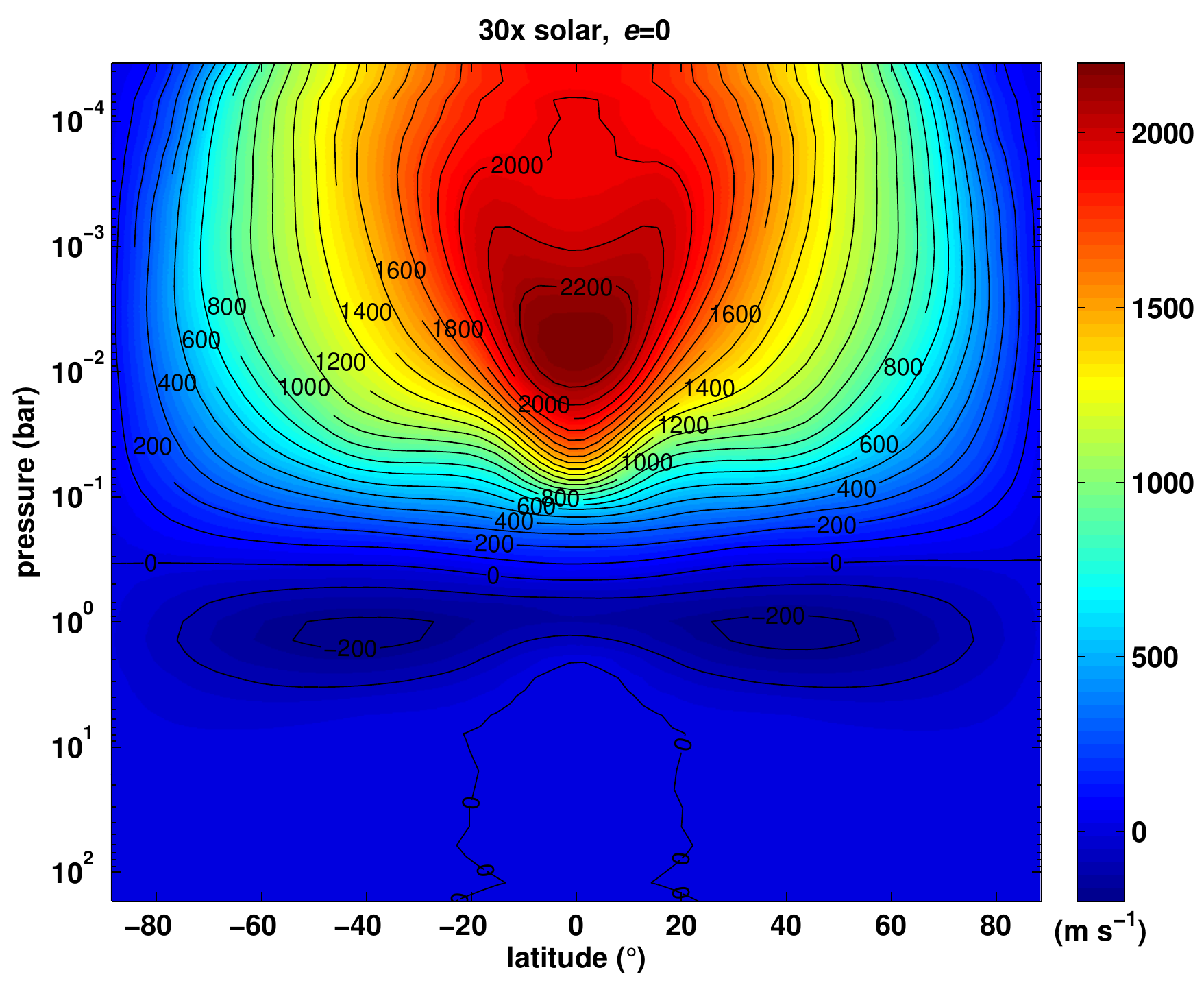}
\caption{Zonal-mean zonal winds for the synchronous rotation ($P_{rot}=P_{orb}$) cases at 1$\times$ (left) and 30$\times$ (right) solar 
metallicity GJ436b atmospheres.  The top panels represent the jet structure for a synchronously rotating 
GJ436b in an eccentric orbit, while the bottom panels assume a circular orbit with the same semimajor axis.
The wind speeds presented here represent 100 day averages of the zonal 
winds taken after each simulation was considered to have reached an equilibrium state.
Note that jet structures for the synchronous cases are very similar to the zonal-mean zonal wind plots 
presented in Figure \ref{uave} for the same atmospheric metallicity.(a color version of this figure is available in the online journal.)}\label{uave_synch}
\end{figure}

\begin{figure}
 \centering
   \includegraphics[width=0.49\textwidth]{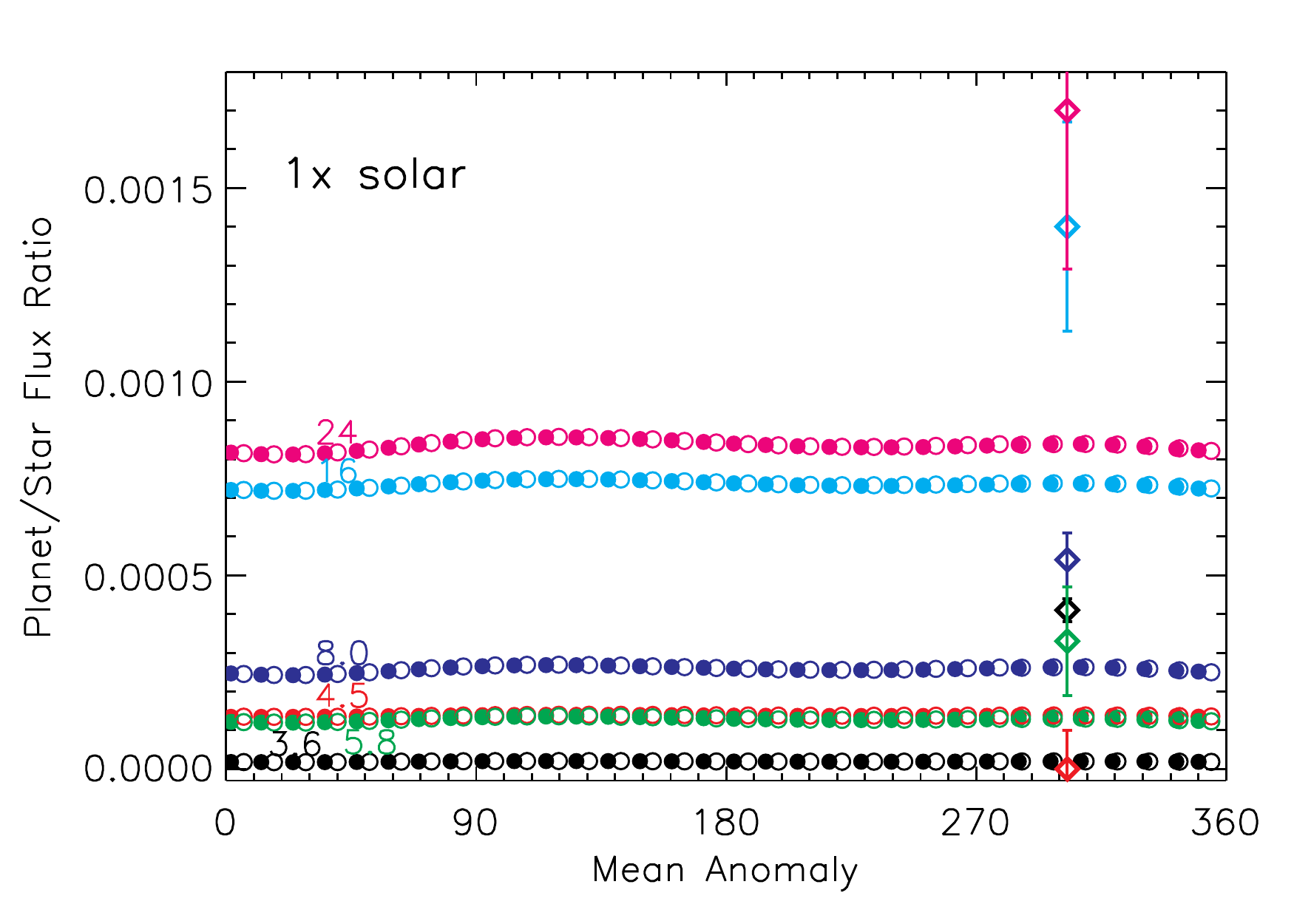}\\
   \includegraphics[width=0.49\textwidth]{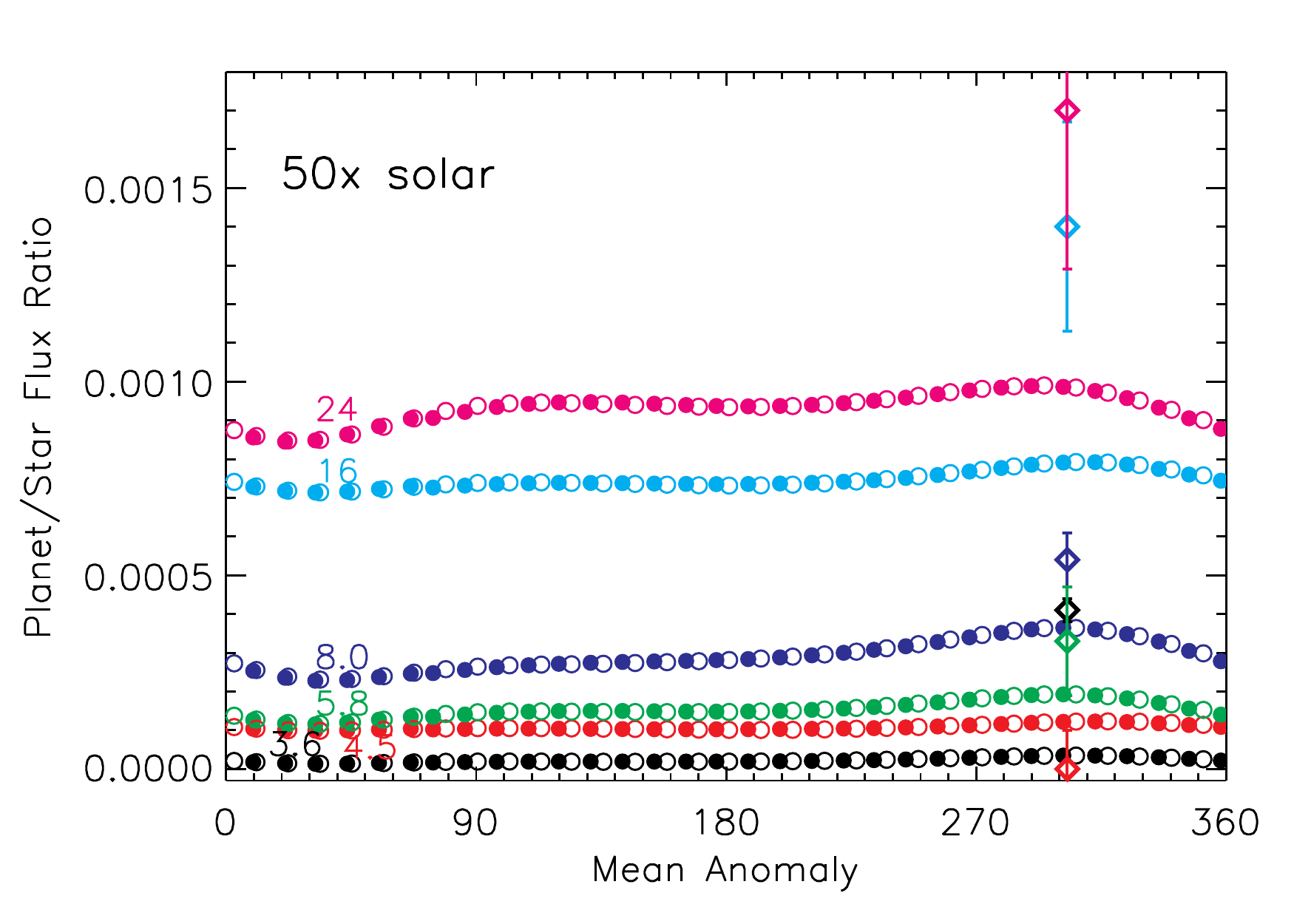}
\caption{Planet/Star flux ratio as a function of orbital phase in each of the {\it Spitzer} bandpasses for the 1$\times$ and 50$\times$ solar 
metallicity cases of GJ436b.  The mean anomaly is an angle that increases linearly with time from periapse passage.  For GJ436b, transit 
and secondary eclipse occur at a mean anomalies of 90$^{\degr}$ and 303$^{\degr}$ respectively.  The filled and open circles represent data 
extracted with 100 simulated days of separation to test for temporal variability, which appears to be minimal.  The secondary eclipse 
measurements from \citet{ste10} are shown for comparison. (a color version of this figure is available in the online journal.)}\label{light_curves}
\end{figure}

The SPARC model is uniquely equipped to produce both theoretical light curves and spectra that account not only for radiative 
effects, but also dynamic movement in the atmosphere.  Once each of the GJ436b models reached an equilibrium state, pressure 
and temperature profiles were recorded along each grid column at many points along the planet's orbit (Figure~\ref{orbit_fig}).  
These pressure-temperature profiles were then used in high resolution spectral calculations to determine the emergent flux from 
each point on the planet and which portion of that emergent flux would be directed toward an earth observer including
limb darkening/brightening effects.  Spectra and light curve generation methods are fully described in \citet{for06}.
Figure \ref{light_curves} shows the theoretical light curves, expressed as the planet/star flux ratio, for the 1$\times$ 
and 50$\times$ solar cases as a function of orbital position.  The 1$\times$ solar light curves are relatively flat due to 
the lack of a day/night temperature contrast as seen in Figure \ref{uvt_plot}.   The day/night temperature contrast is more 
prominent in the 50$\times$ solar case, which results in an increase in the planet/star flux ratio at secondary eclipse.  The 
open and filled circles in Figure \ref{light_curves} represent output from three consecutive orbits separated by 100 
simulated days to test for temporal variability in the light curves.   Little variation is seen in the predicted light 
curves from orbit to orbit or over longer timescales.   

\begin{figure}
 \centering
   \includegraphics[width=0.49\textwidth]{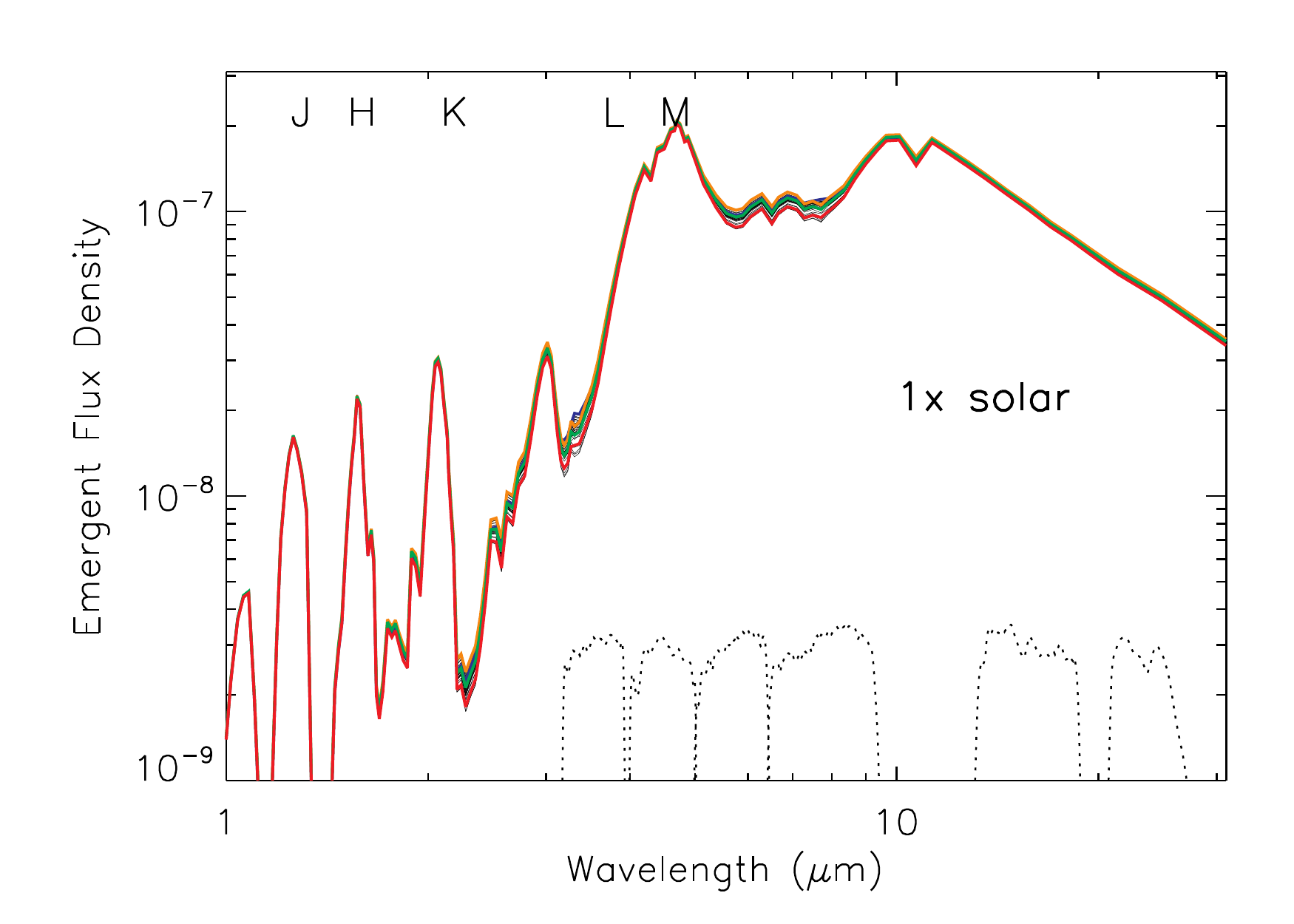}\\
   \includegraphics[width=0.49\textwidth]{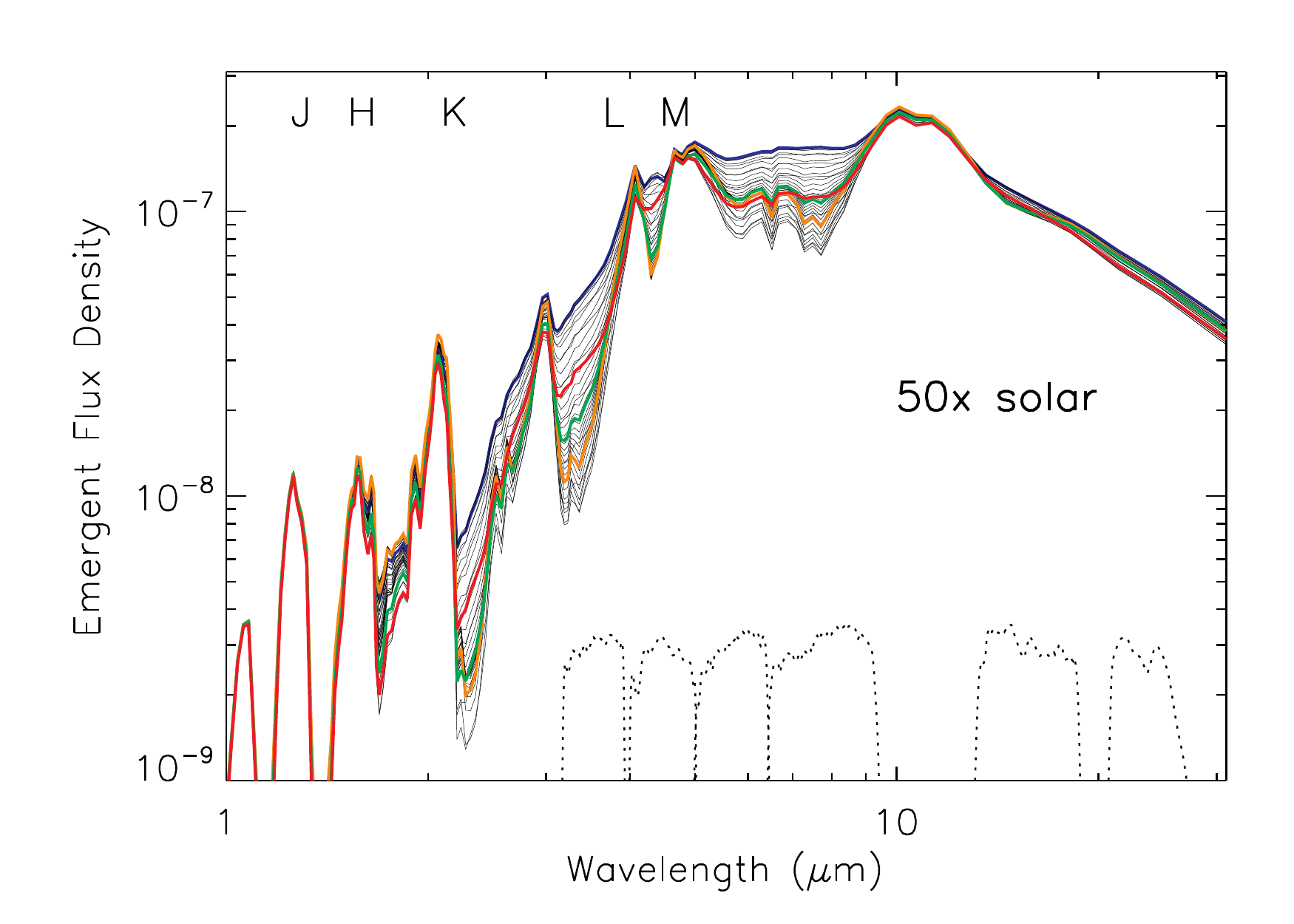}
\caption{Flux per unit frequency, $F_{\nu}$ (erg s$^{-1}$ cm$^{-2}$ Hz$^{-1}$), as a function of wavelength for the 1$\times$ (top) and 50$\times$ (bottom)
solar metallicity cases of GJ436b.  The black spectra presented for each case are taken from 32 locations along a single orbit as shown in Figure 
\ref{orbit_fig}.  The central wavelengths of J-, H-, K-, L-, and M-bandpasses are indicted by the corresponding letters at the top of the plot. 
Dotted lines at the bottom indicate the bandpasses of the four {\it Spitzer} IRAC bands from 3-9 $\mu$m, IRS blue filter at 16 $\mu$m, 
and the MIPS 24 $\mu$m bandpass. Colored spectra are taken from secondary eclipse (blue), 
periapse (red), transit (orange), and apoapse (green). The 50$\times$ solar case shows a strong dependence of the emergent flux density 
on orbital position while the 1$\times$ solar case does not.(a color version of this figure is available in the online journal.)}\label{fluxes}
\end{figure}

Computed emission spectra from the points along the orbit shown in Figure~\ref{orbit_fig} are presented in Figure~\ref{fluxes} 
for both the 1$\times$ and 50$\times$ solar cases. As with all highly irradiated planets, the expectation is that the infrared 
spectra are carved predominantly by the absorption bands of H$_2$O vapor.  These bands are most prominent in the near infrared 
between the J-, H-, and K-band flux peaks (it is at the flux peaks that the H$_2$O opacity is low).  At wavelengths where 
H$_2$O opacity is low, the deeper, hotter layers of the atmosphere can be seen and the emitted flux is generally higher.  However, 
other molecules can imprint absorption features on these emission peaks, and lessen them.  This is true in the 4-5 $\mu$m range
where CO absorbs over the redder half of this range ($\sim$4.5-5 $\mu$m) along with CO$_2$ ($\sim$4.3 $\mu$m), which is 
prominent at high metallicity.  Given the assumption of chemical equilibrium, CH$_4$ is abundant in all of 
the metallicity cases, which leads to a strong absorption band centered on  3.3 $\mu$m, as well as a broad band from 
$\sim$7-8.5 $\mu$m.  However, as the atmospheric metallicity is increased, the CO abundance increases linearly, and the 
CO$_2$ abundance increases quadratically \citep{lod02, zah09}, which leads to a weakening of the CH$_4$ bands and a 
strengthening of the CO and CO$_2$ bands.  In the 50$\times$ solar case a strong CO$_2$ band at 4.4 $\mu$m is clearly 
visible that does not appear in the 1$\times$ solar case.  At high metallicity, this CO$_2$ absorption band 
forces flux out a longer and shorter wavelengths, away from the $\sim$4-5 $\mu$m flux peak.

\begin{figure}
 \centering
   \includegraphics[width=0.49\textwidth]{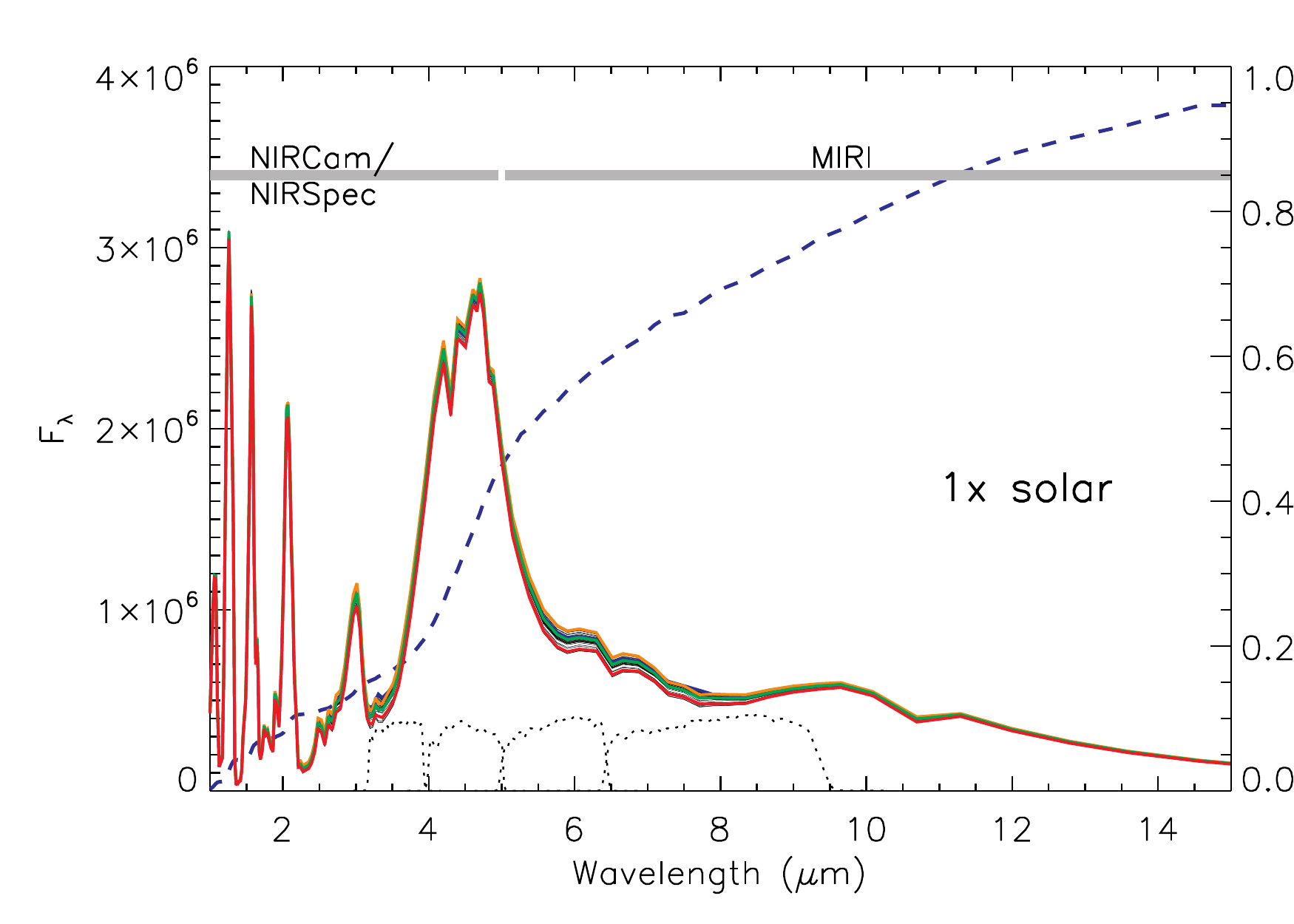}\\
   \includegraphics[width=0.49\textwidth]{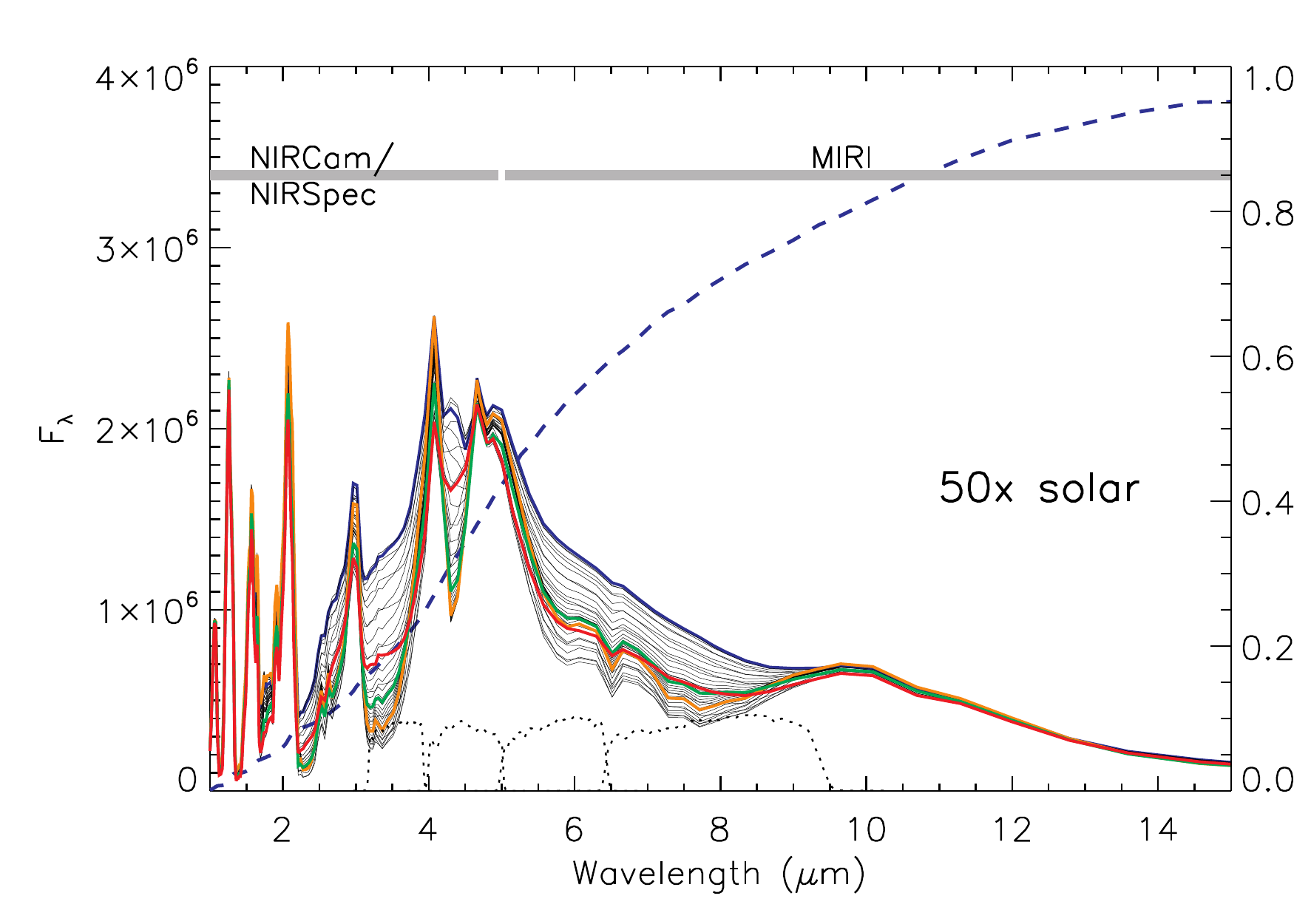}
\caption{Flux per unit wavelength, $F_{\lambda}$ (erg s$^{-1}$ cm$^{-2}$ $\mu$m$^{-1}$), as a function of wavelength for the 1$\times$ (top) and 50$\times$ (bottom) solar 
metallicity cases of GJ436b.  The spectra presented for each case are taken from several points along the 
orbit as shown in Figure \ref{orbit_fig}, with secondary eclipse (blue), periapse (red), transit (orange), and 
apoapse (green) highlighted. The dashed line represents the integrated flux as a function of wavelength from the planet at secondary eclipse and is tied to 
the axes on the right with 1.0 indicating the total integrated flux over all wavelengths.  Dotted lines at the bottom indicate the bandpasses 
of the four {\it Spitzer} IRAC bands from 3-9 $\mu$m.  Solid gray bars at the top indicate the wavelength 
coverage of 3 planned instruments for the JWST:  NIRCam, NIRSpec, and MIRI.  The vertical position of the bars is arbitrary and does not signify 
instrument sensitivity.(a color version of this figure is available in the online journal.)}\label{energy}
\end{figure}

In addition to Figure \ref{fluxes} showing emission spectra from 1 to 30 $\mu$m, emitted flux distributions as a 
function of wavelength, with orbital phase, can also be analyzed.  Figure \ref{energy} presents the planetary flux 
per unit wavelength for the 1$\times$ and 50$\times$ solar cases.
In both the 1$\times$ and 50$\times$ solar models, the peak energy output of the planet occurs between 4 and 5 $\mu$m.  
For the 50$\times$ model, the strong CO and CO$_2$ bands just shortward of 5 $\mu$m dramatically decrease the energy output there, 
forcing much of the energy out at both longer and shorter wavelengths.  The dashed line 
in Figure \ref{energy} shows the integrated flux as a function of wavelength for secondary eclipse (blue dashed line, Figure~\ref{energy}). 
The flux from the planet in the 1-15 $\mu$m range accounts for 95\% of the planet's emergent energy, which makes this an especially 
important wavelength range for determining the atmospheric properties of GJ436b. 

It is interesting to note the increased variability in the emitted flux from the planet as a function of orbital position in the 
50$\times$ solar case compared with the 1$\times$ solar case in Figures~\ref{fluxes} and \ref{energy}.  The flux emitted from the 50$\times$ 
solar case at secondary eclipse (blue line, Figure \ref{fluxes}) is lacking in many of the predominant spectral features seen at other orbital phases, 
due to a shallower day-side temperature gradient.    
The absorption features due to CH$_4$ are much weaker at secondary eclipse indicating a reduction in CH$_4$ abundance 
on the day-side compared to the night-side which is seen during transit (orange line, Figure \ref{fluxes}).  Observing the flux emitted 
from GJ436b as a function of wavelength at several different points along its orbit could reveal a great deal about its overall chemical composition.

\section{Discussion}\label{dis}

The atmospheric models presented here are not only useful for
exploring circulation regimes, chemistry, and radiative transfer, but can also provide insight into
current observations and help guide future observations of GJ436b. Figure~\ref{light_curves} also 
includes the available {\it Spitzer} secondary eclipse measurements from \citet{ste10}.
In all metallicity cases, our predicted planet/star flux ratio falls short of the measured 3.6, 5.8, 
8.0, 16.0, and 24.0 $\mu$m values and is higher than the observed 4.5 $\mu$m value.  However, it
is useful to note that 50$\times$ solar model planet/star flux ratio comes much closer to matching the 
observations than the 1$\times$ solar model, which could hint at a high metallicity as already suggested 
by \citet{ste10}.  High metallicity solutions ($\sim$30$\times$ solar) are also favored by \citet{spi10} 
to match the 8 $\mu$m observations of GJ436b from \citet{demi07}.
The predicted planet/star flux ratios in the 50 $\times$ solar case are within two 
sigma of the values measured by \citet{ste10} at all bandpasses except 3.6 $\mu$m.

In comparing our predicted planet/star flux ratios with those observed by \citet{ste10} it is important to remember
that we have not altered the temperature or chemistry in our models in an attempt to match observations.  The interplay
between the equilibrium chemistry mixing ratios, the absorption and emission of flux, and the atmospheric dynamics dictates
the temperature structure and emergent spectrum.  As pointed out in \citet{ste10} it is likely that disequilibrium chemistry
plays a strong role in the atmosphere of GJ436b. They suggest a CO/CH$_4$ ratio that is many orders of magnitude larger than what
one would predict from an equilibrium chemistry model.  Enhancement of CO at the expense of CH$_4$ is well known in the 
atmospheres of the solar system's giant planets and brown dwarfs, but not to such an extreme degree.
Additionally, they suggest that the photochemical destruction of CH$_4$ is important to further lessen this molecule's importance
in the planet's atmosphere.  The carbon chemistry of an atmosphere will strongly affect flux measurements in the 3.6, 4.5, 5.8,
and 8.0 $\mu$m bandpasses.  As shown in Figure~\ref{fluxes} and discussed in Section~\ref{spec}, lowering the amount
of CH$_4$ in the atmosphere will increase the emergent flux from the planet in the 3.6 and 8.0 $\mu$m bandpasses.  Higher order
hydrocarbons produced by mixing and photochemistry \citep{zah09_2}, such as C$_2$H$_2$, C$_2$H$_4$, and C$_2$H$_6$, are also strong absorbers
throughout the near- and mid-infrared.  Increasing the amount of CO$_2$, along with CO, in the atmosphere will decrease the
emergent flux from the planet in the 4.5 $\mu$m bandpass while at the same time increasing the emergent flux at wavelength on
either side of this bandpass, including the 3.6 and 5.8 $\mu$m bandpasses.

\begin{figure}
 \centering
   \includegraphics[width=0.55\textwidth]{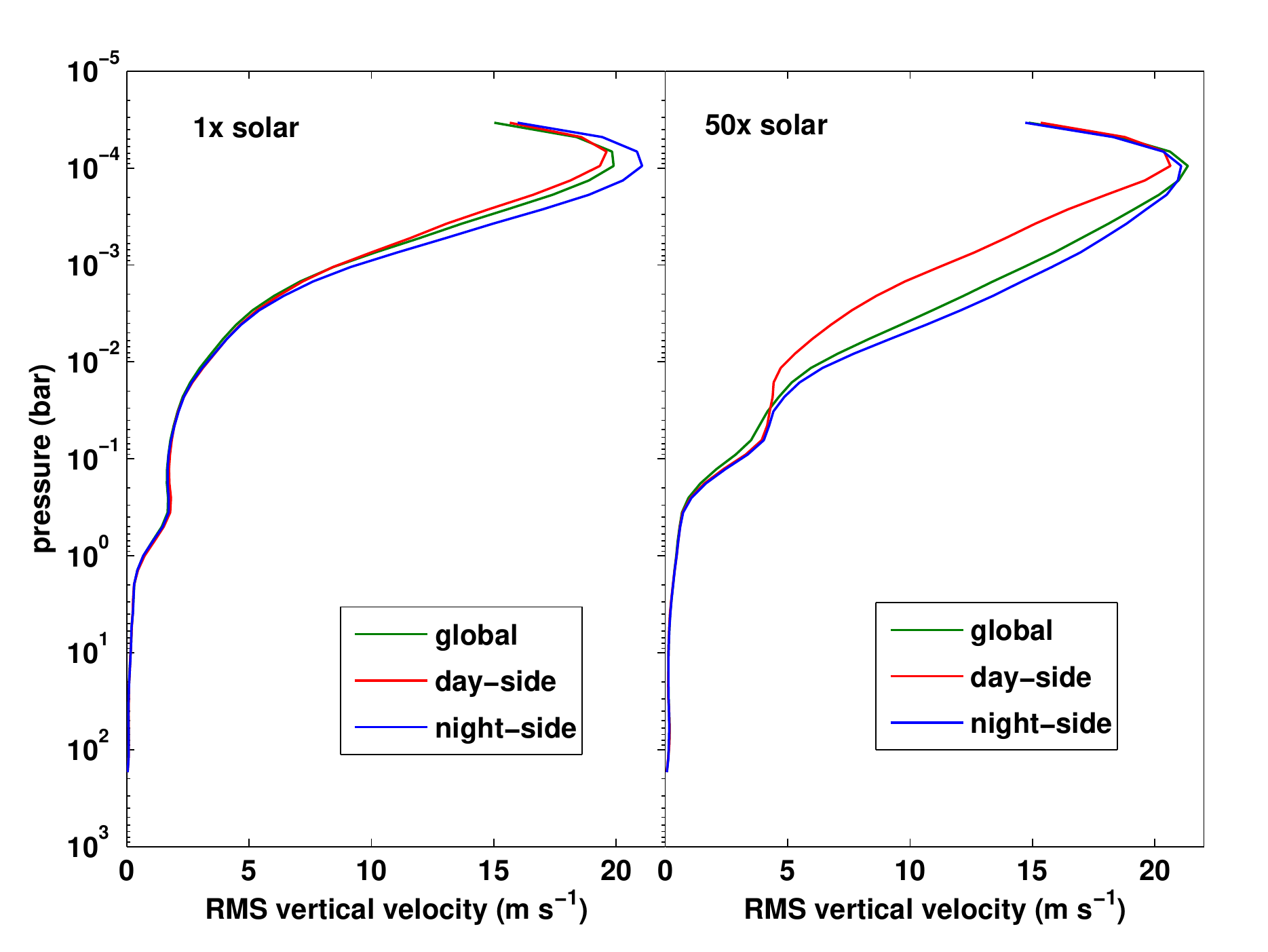}
\caption{RMS vertical velocity ($w_{RMS}$) as a function of pressure for both the 1$\times$ (left) and 50$\times$ solar metallicity cases of GJ436b 
at secondary eclipse.  Global (green line), day-side (red line), and night-side (blue line) averages of the vertical velocity are 
presented.  These $w_{RMS}$ profiles are useful in determining the vertical mixing rate in the atmosphere as a function of pressure 
for use in disequilibrium chemistry and photochemical models. Note that the decrease in $w_{RMS}$ near the top of the domain results 
from the existence of the model's upper boundary where $w$ is forced to be zero.  The vertical velocities would likely
continue increasing with altitude if the top of the model had been  placed at even lower pressures.
(a color version of this figure is available in the online journal.)}\label{vert_vel}
\end{figure}

Although our models do not include disequilibrium chemistry, they can provide an important constraint for disequilibrium
chemistry models---namely, estimates of dynamical timescales and vertical mixing rates.   Disequilibrium
chemistry is expected in all regions of the atmosphere where dynamic timescales, $\tau_{dyn}$, are shorter than
chemical timescales, $\tau_{chem}$.  The dynamic timescale is given simply by
\begin{equation}
\tau_{dyn}=\frac{L}{V}
\end{equation}
where $L$ is the relevant length scale in the horizontal or vertical direction and $V$ is horizontal or vertical wind speed.
For one-dimensional photochemical models, the timescales considered are only those in the vertical direction in which case
$V=w$, where $w$ is the vertical velocity as a function of height or pressure in the atmosphere.  Given the values of
$\omega=Dp/Dt$ from our models, the vertical velocity can be estimated as
\begin{equation}
w=\frac{-H}{p}\omega
\end{equation}
where $H$ is the scale height of the atmosphere, which is a function of $p$ the pressure at given atmospheric level.
Figure~\ref{vert_vel} shows the RMS values for $w$ calculated as global, day-side, and night-side averages for
both the 1$\times$ and 50$\times$ solar cases at secondary eclipse.  The RMS values for $w$
were calculated as $w_{RMS}(p) = \sqrt{A^{-1} \int w^2\,dA}$, where the integral is a horizontal integral (at constant pressure) over the
day-side, night-side, or entire globe as appropriate.  In both the 1$\times$ and 50$\times$ solar cases
the vertical wind speeds increase monotonically from pressures around 1 bar to 0.1 mbar with peak speeds
around $22\rm\, m \,s^{-1}$.  The vertical eddy diffusion coefficient, $K_{zz}$, can be calculated from these vertical
wind speed profiles simply by multiplying by the relevant vertical length scale, $L$.  As shown in \citet{smi98}, in order 
to calculate the correct value of $L$ at each pressure level both dynamic and chemical timescales must
be considered, but to first order $L=H$ especially near the quench level where $\tau_{dyn}=\tau_{chem}$.
Since $H\sim300$ km for GJ436b, one can expect $K_{zz}$ values from 10$^{8}$ cm$^2$ s$^{-1}$ near 100 bar to
10$^{11}$ cm$^2$ s$^{-1}$ near 0.1 mbar.  These values of $K_{zz}$ are in line with those favored by \citet{mad10} 
to explain the disequilibrium CO/CH$_4$ ratio needed to fit the \citet{ste10} observations.

Regardless of the detailed chemistry, our models suggest that
the basic circulation regime of planets in the temperature range of GJ436b
depend strongly on overall metallicity.  High metallicity
cases (30--50$\times$ solar) produce a dominant eastward equatorial jet
that peaks on or near the equator (Figure~\ref{uave}).
This pattern resembles that previously obtained for more strongly irradiated
hot Jupiters \citep{coo05, sho08, sho09, rau10}.  At low metallicity, on the
other hand, the equatorial jet weakens and fast eastward jets develop in midlatitudes---a 
jet pattern distinct from those previously reported in the hot-Jupiter modeling
literature.  These changes suggest that the atmosphere experiences a regime
shift where different dynamical mechanisms control the jet structure 
at low versus high metallicity.  It may be that each regime is relevant
to a range of planets of differing effective temperatures and compositions,
providing a motivation to better understand their underlying dynamics.
We will discuss the specific mechanisms responsible for this regime shift
in a future paper.

The atmospheric circulation models presented here represent a first
step in better understanding the dynamical, radiative, and
chemical mechanisms that shape the atmosphere of GJ436b.   Although
chemical disequilibrium is likely to play an
important role, it is important to first consider ---as we have done here--- a baseline
atmospheric model where the effects of chemical equilibrium opacity
are tested before more complex chemistry, clouds, and other factors
are added and complicate the interpretation of the results.
It is unlikely that the presence of disequilibrium chemistry in the atmosphere of 
GJ436b would strongly affect the global circulation patterns seen in this study.
\citet{mad10} suggest that a high metallicity ($\sim$30$\times$ solar)
atmosphere, with chemical disequilibrium induced both by thermal
quenching and photochemistry, can best explain the observations of
\citet{ste10}.  If this is the case, then one would
expect global circulation patterns on GJ436b akin to our 30$\times$
and 50$\times$ solar models with a strong equatorial
jet and non-negligible day/night temperature contrasts.  As shown in
Figures \ref{light_curves} and \ref{fluxes} one
would expect to see variations in the planet/star flux ratio as a
function of orbital phase for a high metallicity
GJ436b-like atmosphere.  This variation in the planet/star flux ratio
could reveal a great deal about circulation patterns on
GJ436b along with day/night chemistry differences.  For this reason we
recommend that full-orbit light curves of the planet in
the 3.6 $\mu$m bandpass be obtained during the Warm {\it Spitzer}
mission. Looking further into the future, with the {\it James Webb Space Telescope} (JWST) a
combination of NIRSpec or NIRCam data from 3-5 $\mu$m, as well as MIRI
data shortward of $\sim$15 $\mu$m, will sample the vast majority of
the planet's emitted energy. Observations as a function of orbital
phase with these instruments could tightly constrain the energy budget
of the planet.  Detection of the CO$_2$ band depth at 4.4 $\mu$m
with NIRSpec would be also an important metallicity indicator and
probe of the carbon chemistry of GJ436b.

\section{Conclusions}\label{conc}

In this study we have investigated atmospheric circulation for the hot Neptune GJ436b for various assumed atmospheric metallicities
using a three-dimensional coupled radiative transfer and general circulation model.  We have found that assumed atmospheric composition 
for a planet like GJ436b can have strong effects on both day/night recirculation and jet patterns within the atmosphere.  Light curves 
and spectra produced from our models show that enhancements in atmospheric metallicity over solar produce an increase as well as 
orbital phase variations in the planet/star flux ratio.  Given the possible strong dependence of emitted 
flux with orbital phase, GJ436b could provide an important probe of atmospheric chemistry outside of our solar system.  
Moreover, we showed that for warm gaseous extrasolar planets in the effective temperature range of $\sim$600-900 K, there exists a regime 
shift as metallicity is increased from a circulation dominated by mid-latitude jets and minimal longitudinal temperature differences 
to one dominated by an equatorial jet and large day-night temperature differences.
Although these models of GJ436b do not include all the processes at play in any given atmosphere, the basic trends for 
atmospheric circulation as a function of metallicity will be useful in better understanding GJ436b and many other extrasolar 
planetary atmospheres.

\acknowledgments

This work was supported by NASA Headquarters under the NASA Earth and Space Science 
Fellowship Program (Grant NNX08AX02H) and Origins Program (Grant NNX08AF27G).  
Work by KL was supported by NSF grant AST 0707377 and by the National Science Foundation, while working at the Foundation.
Resources supporting this work were provided by the NASA High-End Computing (HEC) Program through the NASA 
Advanced Supercomputing (NAS) Division at Ames Research Center.  The authors 
would like to thank Yuan Lian for his helpful insights into working with 
the MITgcm and the anonymous referee for their helpful comments and suggestions.

\bibliographystyle{apj}
\bibliography{ms_astroph}

\clearpage

\onecolumn

\clearpage

\clearpage

\clearpage

\clearpage

\clearpage

\clearpage

\clearpage

\clearpage

\clearpage

\clearpage

\clearpage

\clearpage

\clearpage

\begin{table}
\caption{GJ436A/b parameters.}\label{gj436_params}
\centering
\begin{tabular}{rc}
\tableline\tableline
Parameter & Value \\
\tableline
$R_p$ (R$_J$) & 0.3767 \\
$M_p$ (M$_J$) & 0.0729 \\
$g$ (m s$^{-2}$) & 12.79 \\
$a$ (AU) & 0.02872\\
$e$ & 0.15\\
$\varpi$ (deg) & 343\\
$P_{orb}$ (days) & 2.64385 \\
$P_{rot}$ (days) & 2.32851 \\
$R_{\star}$ (R$_{\odot}$) & 0.464\\
$M_{\star}$ (M$_{\odot}$) & 0.452\\
$T_{eff}$ (K) & 3350\\
\tableline
\end{tabular}
\tablecomments{Planetary and stellar parameters taken from \citet{tor08}. Values for $e$ and $\varpi$ were 
taken from \citet{demi07}.}
\end{table}

\end{document}